\documentclass{osa-article}

\usepackage[utf8]{inputenc}
\usepackage{graphicx}
\usepackage{amsmath}
\usepackage{amssymb}
\usepackage{subcaption}
\usepackage{enumitem}
\usepackage{appendix}
\usepackage{theorem}
\setlist{leftmargin=*}

\newtheorem{theorem}{Theorem}

\journal{oe}

\articletype{Research Article}

\begin{document}

\title{On Space-Spectrum Uncertainty Analysis for Coded Aperture Systems}

\author{Vishwanath Saragadam and Aswin~C.~Sankaranarayanan}

\address{ECE Department, Carnegie Mellon University, Pittsburgh, PA, USA}

\email{ vishwanathsrv@cmu.edu; saswin@andrew.cmu.edu} 



\begin{abstract}
We introduce and analyze the concept of space-spectrum uncertainty for certain commonly-used designs for spectrally programmable cameras.
Our key finding states that, it is impossible to simultaneously capture high-resolution spatial images while programming the spectrum at high resolution.
This phenomenon  arises due to a  Fourier relationship between the aperture used for obtaining spectrum and its corresponding diffraction blur in the (spatial) image.
We show that the product of spatial and spectral standard deviations is lower bounded by $\frac{\lambda}{4\pi \nu_0}$ femto square-meters,
where $\nu_0$ is the density of groves in the diffraction grating and $\lambda$ is the wavelength of light.
Experiments with a lab prototype for simultaneously measuring spectrum and image validate our findings and its implication for spectral filtering.
\end{abstract}

\section{Introduction}\label{section:intro}
Spectrum is often a unique feature of materials and is used for identification and classification across a broad range of fields such as geology \cite{cloutis1996review}, biological imaging \cite{colthup2012introduction,lichtman2005fluorescence} and material identification \cite{saragadam2019programmable,zhi2019multispectral}.
Tools such as the hyperspectral camera capture the spectrum of a scene which is subsequently used for identification and classification purposes.

Capturing the full spectrum, while useful, is also quite wasteful especially if we are only interested in measuring similarity of the spectral profile at each pixel to a small collection of reference spectra.
Such cameras, called \textit{spectrally-programmable cameras}, have been demonstrated \cite{mohan2008agile,love2014full} with compelling applications in computer vision.
This paper evaluates a spectrally programmable cameras involving a dispersive element, such as a diffraction grating or a prism, and a spatial light modulator (SLM).

\begin{figure}[!tt]
	\centering
	\includegraphics[width=\textwidth]{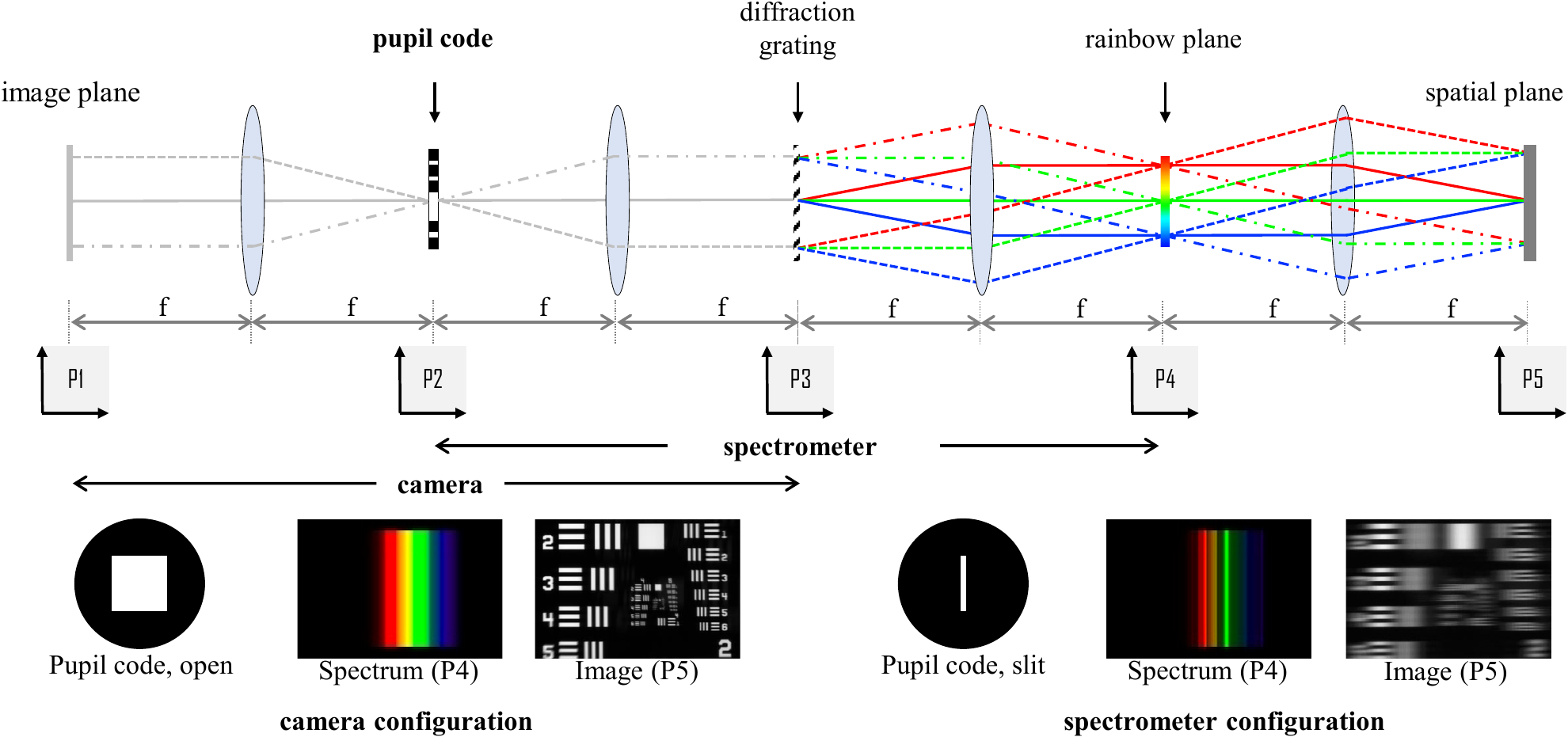}
	\caption{\small \textbf{Optical setup for capturing images with spectral modulation}. P1 is the image plane of the objective lens, P2 contains spatial frequencies of the image, where we place a pupil code, $a(x, y)$. P3 contains the image plane blurred by the aperture function. We place a diffraction grating at this plane to disperse light into different wavelengths. P4 contains the resultant spectrum and P5 is a flipped copy of P3. In a camera configuration, the pupil code consists of an open aperture and leads to sharp image but blurred spectrum. In spectrometer configuration, the pupil code is a slit and leads to sharp spectrum but blurred image. Our paper formalizes the role played by pupil code for spatial and spectral resolutions.}
	\label{fig:schematic}
\end{figure}

\subsection{Problem setting} 
The analysis in this paper is for the optical setup shown in Fig. \ref{fig:schematic}, commonly used in prior art for spectral filtering \cite{love2014full,lin2014dual,saragadam2018krism,saragadam2019programmable}.
The optical system consists of a series of lenses of focal length $f$, each subsequent pair separated by $2f$. 
The  setup relays the image plane from plane P1 to P3 with a \textit{pupil code} in P2.
A diffraction grating placed on P3 provides a spectral dispersion of the light.
The dispersed light is focused on plane P4 to form the so-called \emph{rainbow plane}, where each point corresponds to the average intensity of light of the whole scene for a single wavelength. 
The image on P3 is simply relayed on to plane P5.
Arbitrary spectral filtering can then be performed by placing an (SLM) on the rainbow plane (P4) and measuring image on plane P5.
Intuitively, planes P1 to P3 is a simple camera with an aperture in its Fourier plane, and planes P2 to P4 is a spectrometer with the aperture replacing a slit.
The setup provides an insight into the tradeoff between spatial and spectral resolutions. While a camera requires a large and open aperture for a compact spatial blur, this would lead to severe loss in spectral resolvability, as a spectrometer requires a narrow opening. 
Our goal is to formalize the role played by the shape of the pupil code in deciding spatial and spectral resolution.

\subsection{Main Result}
We show that the pupil code $a(x, y)$ introduces a spectral and spatial blur, $h_\lambda (\lambda)$ and $h_x(x)$ respectively with standard deviations $\sigma_\lambda$ and $\sigma_x$ (detail expressions in eq. \eqref{eq:spreads}).
Our main contribution is in the form of a \emph{space-spectrum bandwidth product} that  relates the spectral resolution at which light can be modulated and the spatial resolution of captured image.
This is encapsulated in the following theorem,
\begin{theorem}
	For the spectrally-coded imaging architecture shown in Fig.\ \ref{fig:schematic}, the product of spatial and spectral standard deviations $\sigma_x$ and $\sigma_\lambda$ respectively is related as
	\begin{align}
	\sigma_x \sigma_\lambda \ge \frac{\lambda}{4\pi \nu_0},
	\end{align}
	where $\nu_0$ is the density of slits in the diffraction grating.
	\label{th:thm1}
\end{theorem}
This result was first explored in \cite{love2014full} and \cite{saragadam2018krism} where they stated that a slit trades off spatial and spectral resolution.
Our paper builds on their results by providing a concise expression for the tradeoff.
We prove that a Gaussian-shaped pupil code achieves the lower bound and leads to most compact spatial blur for a targeted spectral blur.
We also show that for narrowband filtering, the spatial blur is affected by the pupil code as well as the shape of the narrowband filter, and that a slit, a commonly used narrowband filter shape leads to a spectrally-varying spatial blur.
Instead, using a Gaussian-shaped narrowband filter achieves spectrally-independent spatial blur, thereby being the optimal candidate for spectral programming.

\subsection{Implications}
The space-spectrum bandwidth product introduces an uncertainty in spectrally-modulated cameras, stating that one cannot arbitrarily code spectrum at high resolution without loss in spatial resolution. 
We demonstrate the impact of uncertainty by building a spectrally-programmable camera and showing that blocking one of two closely-spaced narrowband sources cannot be done without severe loss in spectral resolution.

\paragraph{Hyperspectral imagers.} Apart from programmable spectral filtering, several hyperspectral imaging architectures~\cite{lin2014dual,saragadam2018krism,august2013compressive} rely on obtaining spectrally-modulated images.
Our findings have a direct implication on such setups, as a key requirement of such setups is to capture high resolution images without sacrificing spectral resolution.
Hence, the space-spectrum bandwidth product can serve as a design guide to carefully choose the pupil code to obtain desired spatial and spectral resolutions.

\paragraph{Spatially-coded cameras.} We note that the analysis in the paper does not apply to many spectral cameras where there is no pupil plane coding. 
Cameras such as the  pushbroom camera and  the coded aperture snapshot spectral imager (CASSI) \cite{wagadarikar2008single,kittle2010multiframe} only perform spatial coding and, as such, are not affected by this result.
Since such systems code space and then measure its spectrally-sheared image, the spatial code only affects the spectral resolution and not the spatial resolution.

\section{Prior Work}\label{section:prior_work}
We start our discussion by talking about capturing images with arbitrary spectral filters and then briefly state its applications. We then state the fundamental tradeoffs based on system parameters.

\paragraph{Measurement model.} Consider a scene's hyperspectral image (HSI) represented by $H(x, y, \lambda)$, where $(x, y)$ represent spatial coordinates and $\lambda$ represents wavelength.
Our goal is to optically obtain a spectrally-filtered image of the scene with a spectral filter.
Specifically, given a spectral filter $f(\lambda)$, our aim is to implement a camera that captures the following grayscale image,
\begin{align}
I(x, y) &= \int_{\lambda} H(x, y, \lambda) f(\lambda) d\lambda.
\end{align}

\paragraph{Applications of spectral filtering.}
The ability of arbitrarily code spectrum enables a wide gamut of applications.
This includes adaptive color displays \cite{mohan2008agile}, programmatically blocking illuminants \cite{love2014full}, and detecting materials \cite{saragadam2019programmable,zhi2019multispectral}.
The key advantage in all these applications is to not measure the complete HSI, but only the desired spectral projections, which leads to fewer measurements and higher signal to noise ratio (SNR).
Such a system can also be used for compressively sensing the complete HSI \cite{saragadam2018krism,lin2014dual} which relies on capturing projection of a scene's HSI on random or designed spectral filters.

\paragraph{Spectrally-programmable camera architecture.} Spectral programming is an ubiquitous technique that is often used in imaging applications, such as Bayer filters for RGB cameras or narrowband spectral filters for fluorescence microscopy \cite{lichtman2005fluorescence}.
Static filters offer arbitrarily high spectral resolution, but are not tailored for applications that require changing filters rapidly; while this can be achieved with filter wheels, the speed of such devices is constrained by the speed at which the filters can be changed.
Electronically tuning filters, in part can be achieved by using a tunable filter \cite{lctfwiki} where liquid crystal (LC) cells are used to obtain a combination of narrowband spectral filters.
LC filters however are typically slow as they require large settling times.

The most practical way of implementing rapidly changeable programmable spectral filters is to rely on the setup shown in Fig \ref{fig:schematic}.
Here, a dispersion element such as a grating or prism is used to create the so-called \emph{rainbow plane} \cite{mohan2008agile} where each point corresponds to intensity of a single wavelength of the whole scene.
By placing a spatial modulator (SLM) on this plane, one can achieve arbitrary spectral coding.
This approach is similar to replacing a sensor in a spectrometer with an SLM, and has been the \emph{defacto} way of spectral coding in some of the past works \cite{lin2014dual,love2014full}.

The SLM-based approach has certain advantages. 
Since SLMs are fast (often in excess of 60fps), one can achieve high frame rates, which is crucial for imaging dynamic scenes, or settings that require several rapidly switching spectral filters.
Two, the system is \emph{potentially} capable of achieving high spectral resolution without sacrificing capture time.
However, as we will see next, a high spectral resolution has a debilitating effect in the form of a severe loss in spatial resolution.
The focus of this paper will be on the fundamental trade-off of spectral and spatial resolutions when using an SLM-based programmable camera.
For brevity, we will henceforth refer to SLM-based programmable camera as just spectrally-programmable camera.

\paragraph{Time-bandwidth product.}
Our main result is based on the time-bandwidth product \cite{GRAMI201641}, which we state here for completeness. 
Let $x(t)$ be a time-domain signal, and let $X(\omega)$ be its continuous-time Fourier transform. 
We define the spread of time-domain and frequency-domain signals as,
\begin{align}
\sigma_t = \frac{\sqrt{\int_t t^2 |x(t)|^2 dt}}{\sqrt{\int_t |x(t)|^2 dt}} \quad \text{and} \quad \sigma_\omega = \frac{\sqrt{\int_\omega \omega^2 |X(\omega)|^2 d\omega}}{\sqrt{\int_\omega |X(\omega)|^2 d\omega}}.\label{eq:timespread}
\end{align}
Then the time-bandwidth product states that,
\begin{align}
\sigma_t \sigma_\omega \ge \frac{1}{4\pi}.
\label{eq:timebandwidth}
\end{align}
As a consequence, one cannot achieve simultaneous localization in time and frequency.
Our result is a translation of the time-bandwidth product applied to spatial and spectral signals which arises Fourier transform property of a thin lens\cite{goodman2005introduction}. Therefore, the blurs in space and spectrum have a scaled Fourier-pair relationship.

\section{Fundamental Limits of Spatial/Spectral Resolution}\label{section:uncertainty}
\paragraph{Space-spectrum tradeoff.} A key component of a spectrometer is a slit that leads to high resolution spectral measurements.
In similar spirit, a programmable camera would also necessitate a narrow slit to ensure that spectrum can be modulated at high resolution.
However, such a narrow slit leads to a severe loss in spatial resolution, since imaging at high resolution requires a large and open aperture.
The authors in \cite{mohan2008agile} note that a large slit leads to loss of spectral resolution, but they do not mention what happens to the spatial resolution.
The authors in \cite{love2014full} identified this tradeoff and stated an approximate relationship between spectral and spatial tradeoff for fully open aperture and demonstrated that high spectral resolution lead to blurry images.
We formalize the result and show that such a tradeoff applies to any pupil code shape and can be concisely stated as a space-spectrum bandwidth product.
In the upcoming sections, we will formalize the spatial and spectral resolutions that result from the choice of a pupil code.

\paragraph{Spectral and spatial blurs.}
We present the spatial and spectral blurs here and refer the interested readers to Appendix \ref{section:appendix1} for a detailed derivation. 
For brevity, we show the blur along $x$-axis alone, as there is no spectral dispersion along $y$-axis.
Let $a(x)$ be the shape of the aperture function and let $A(\cdot)$ be its Fourier transform.
Then the spectral and spatial blur functions are,
\begin{align}
h_\lambda (\lambda) = |a(-\lambda f \nu_0)|^2, \qquad 
h_x (x) = \left|A\left(-\frac{x}{\lambda f}\right)\right|^2.
\end{align}
We observe that the blur in space and spectrum are not independent; in fact, they behave similar to a Fourier-transform pair, with appropriate scaling.
Our goal is to show that this interdependence between spatial and spectral blur has a very specific structure and can be lower bounded -- implying that we cannot arbitrarily resolve in both domains.

\subsection{The space-spectrum uncertainty principle}
Our main result, stated in Theorem \ref{th:thm1}, suggests that the spatial and standard deviations are related by the inequality,	$\sigma_x \sigma_\lambda \ge \frac{\lambda}{4\pi \nu_0}$.
We now outline the proof of our theorem.
\paragraph{Proof.} We define the spectral and spatial standard deviations as,
\begin{align}
\sigma_\lambda &= \sqrt{\frac{\int_\lambda \lambda^2 h_\lambda (\lambda) d\lambda}{\int_\lambda h_\lambda(\lambda) d\lambda}}
= \sqrt{\frac{\int_\lambda \lambda^2 |a(-\lambda f \nu_0)|^2 d\lambda}{\int_\lambda |a(-\lambda f \nu_0)|^2 d\lambda}}\\
\sigma_x &= \sqrt{\frac{\int_x x^2 h_x(x) dx}{\int_x h_x(x) dx}}
= \sqrt{\frac{\int_x x^2 \left| A\left(-\frac{x}{\lambda f}\right)\right|^2 dx}{\int_x \left| A\left(-\frac{x}{\lambda f}\right)\right|^2 dx}},	
\label{eq:spreads}		   
\end{align}
which are similar to time and frequency spreads defined in eq. \eqref{eq:timespread} with appropriate scaling.
Given that $\sigma_t$ is the spread of $x(t)$, the spread of a scaled function $\widehat{x}(t) = x(st)$ is $\widehat{\sigma}_t = s \sigma_t$.
From eq. \eqref{eq:timebandwidth} and substituting $t = \lambda f \nu_0$ and $\omega = \frac{x}{f \lambda}$,
\begin{align}
\left(\frac{1}{f^2 \lambda^2}\right) \sigma^2_x (f^2 \nu_0^2)\sigma^2_\lambda &\ge \frac{1}{16\pi^2} \implies 
\sigma^2_x \sigma^2_\lambda \ge \frac{\lambda^2}{16\pi^2\nu_0^2}\\
&\boxed{\sigma_x \sigma_\lambda \ge \frac{\lambda}{4\pi \nu_0} } 	 \label{eq:main} 
\end{align}

\paragraph{Implication.} We make some observations about the uncertainty principle here.
\begin{itemize}
	\item \emph{Invariance to scaling.} The bandwidth product does not change even if the aperture is stretched or squeezed. If the aperture $a(x)$ is replaced by $a(sx)$, then spectral blur changes to $h_\lambda (\lambda) = |a(-s\lambda f \nu_0)|^2$ and the spatial blur changes to $\left|A\left(-\frac{x}{s\lambda f}\right)\right|^2$. This changes the spectral and spatial variances to $s^2 \sigma^2_\lambda$ and $\sigma^2_x/s^2$, thereby keeping the product a constant. 
	
	\item \emph{Invariance to power of lenses.} The bandwidth product is independent of focal length of the system, implying that one cannot expect any increase in product of standard deviations by changing the lenses.
	
	\item \emph{Dependence on groove density.} The bandwidth product is a function of groove density $\nu_0$, and decreases with increasing density.
	In theory, one can achieve arbitrarily low space-bandwidth product by having high groove density, but the limiting factor becomes the aperture size of lenses.
	
	\item \emph{Dependence on wavelength.} The bandwidth product is directly proportional to wavelength. 
	This is not a surprising result. The limiting case of our statement, where $\sigma_\lambda$ is several hundreds of nanometers is just a normal grayscale imager, and in that case, the expression looks very similar to Abbe's diffraction limit \cite{opticalphysics1998}.
	However, one may make the expression independent of wavelength by using lower bound of the spectral range,
	\begin{align}
	\boxed{\sigma_x \sigma_\lambda \ge \frac{\lambda_\text{min}}{4\pi \nu_0}}
	\end{align}
\end{itemize}

\begin{figure}[!tt]
	\centering
	\includegraphics[width=\columnwidth]{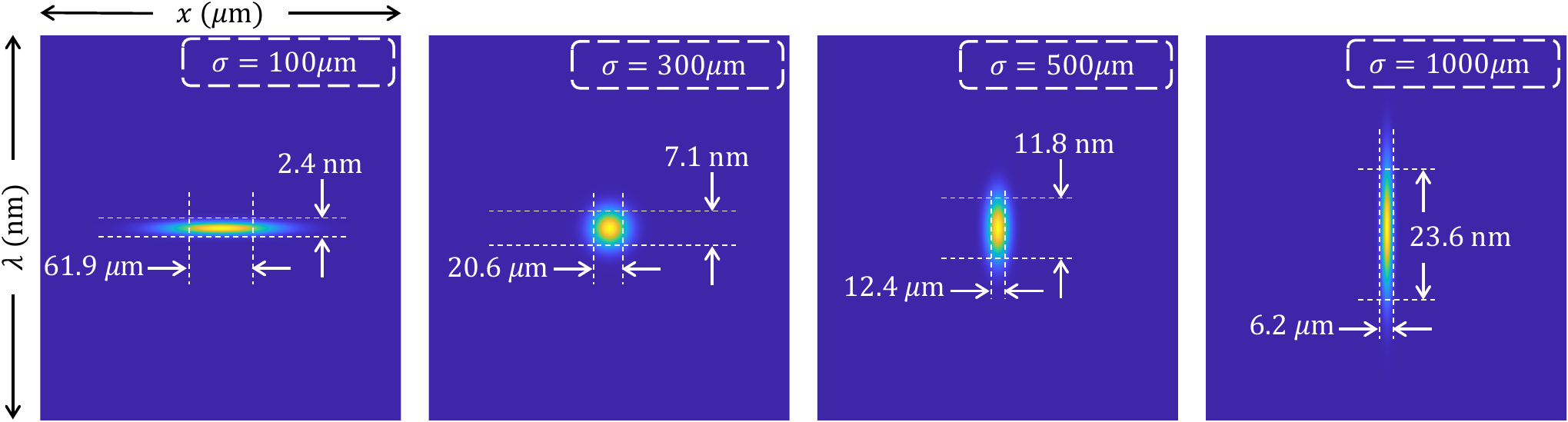}
	\caption{\small \textbf{$\mathbf{x-\lambda}$ blur for Gaussian window of various sizes.} The four figures illustrate the spatio-spectral blur for various window sizes. The blur kernel was computed for $\lambda=500$ nm, $f=100$ mm and a groove density of $300$ grooves/mm. There is a visible trade off between the two resolutions. The appropriate window size depends on the application; a camera with low spectral resolution requirement can use a $\sigma=500 \mu$m window, while one with stringent spatial resolution requirements may use a $\sigma=1000 \mu$m window.}
	\label{fig:xlambda}
\end{figure}

\paragraph{Achievability of lower bound.} As in the case of time-frequency uncertainty, there exists a pupil code function that has its space-spectrum bandwidth product \emph{equal} to $\frac{\lambda}{4\pi \nu_0}$.
This is achieved by the family of Gaussian windows:
\begin{align}
a(x, y) = \text{exp}\left\{\frac{-x^2}{2\sigma^2}\right\}.
\end{align}
The spectral and spatial blur are then given by,
\begin{align}
\widetilde{f}(\lambda) = \text{exp}\left\{-\frac{\lambda^2f^2\nu_0^2}{\sigma^2}\right\},\qquad
\widetilde{g}(x) = \text{exp}\left\{-\frac{4\pi^2\sigma^2x^2}{\lambda^2f^2}\right\}.
\end{align}
Figure \ref{fig:xlambda} shows the simulated ``uncertainty" box  at $500$ nm of Gaussian windows of various widths.
We simulated a system comprising of $75$ mm lenses and a diffraction grating with a groove density of $300$ grooves/mm.
Evidently, as we squeeze along one axis, the other axis stretches with the product of widths being a constant at $145.$9 nm$\cdot\mu$m.
Next, we validate our findings with an optical setup that implements the schematic in Fig.\ \ref{fig:schematic} and capture scenes with various aperture shapes.

\section{Experiments}\label{section:real}

\begin{figure}[!tt]
	\centering
	\begin{subfigure}[c]{0.4\columnwidth}
		\centering
		\includegraphics[width=\columnwidth]{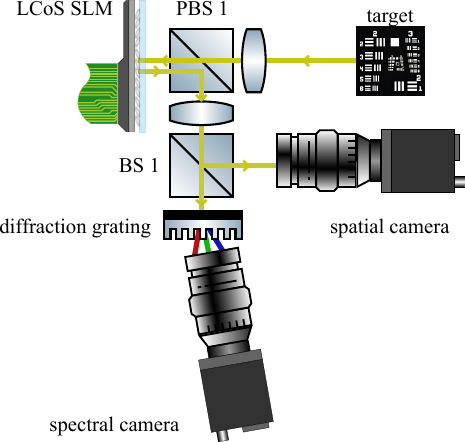}
		\caption{Schematic}
	\end{subfigure}
	\qquad
	\begin{subfigure}[c]{0.4\columnwidth}
		\centering
		\includegraphics[width=\columnwidth]{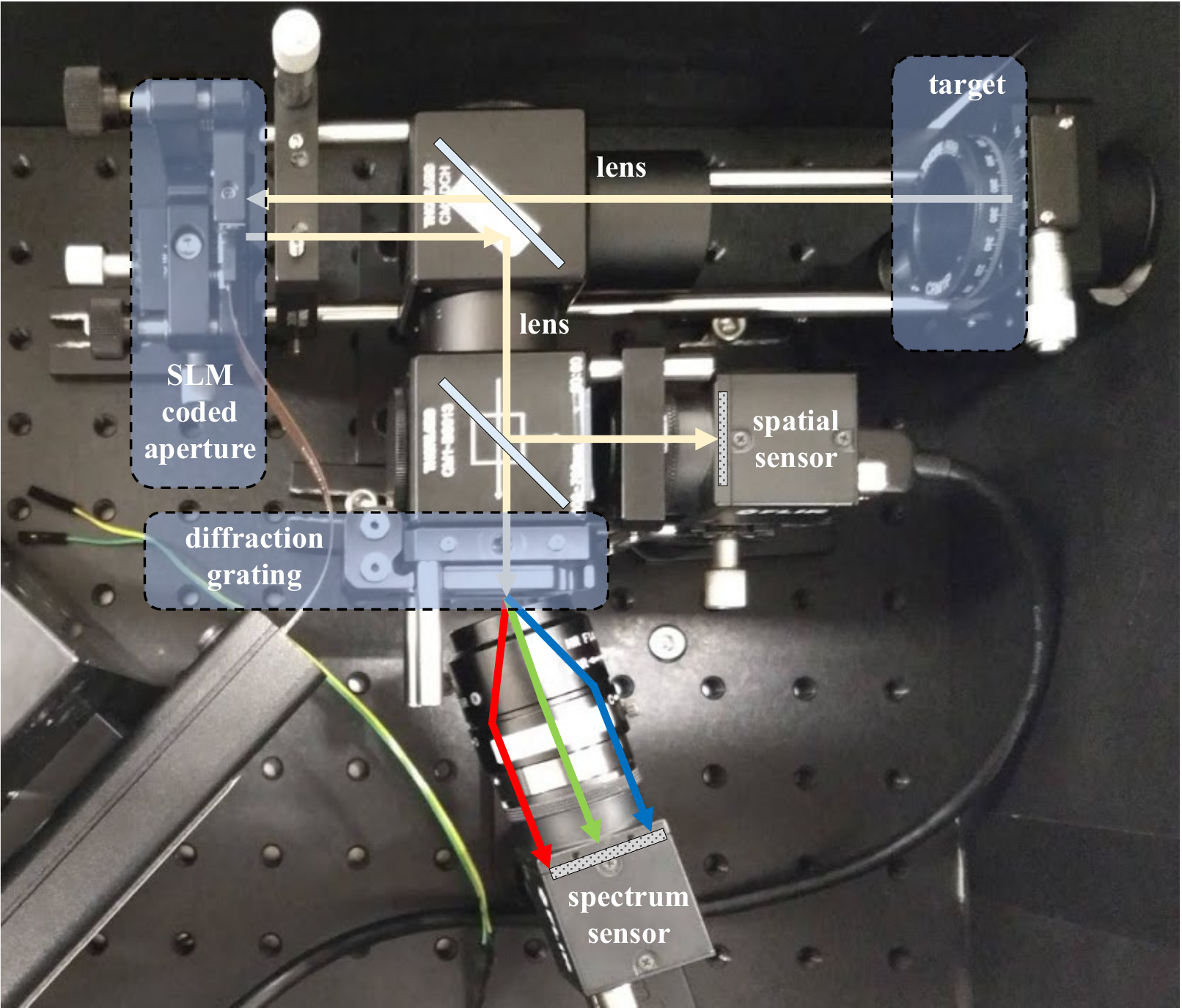}
		\caption{Lab prototype}
	\end{subfigure}
	\caption{\small \textbf{Schematic and image of our lab prototype}. We displayed various patterns on SLM to form coded apertures to evaluate spatial and spectral resolutions. The spectral camera was tilted to capture the first order of diffraction from the grating.}
	\label{fig:setup}
\end{figure}
\begin{figure}[!tt]
	\centering
	\includegraphics[width=\columnwidth]{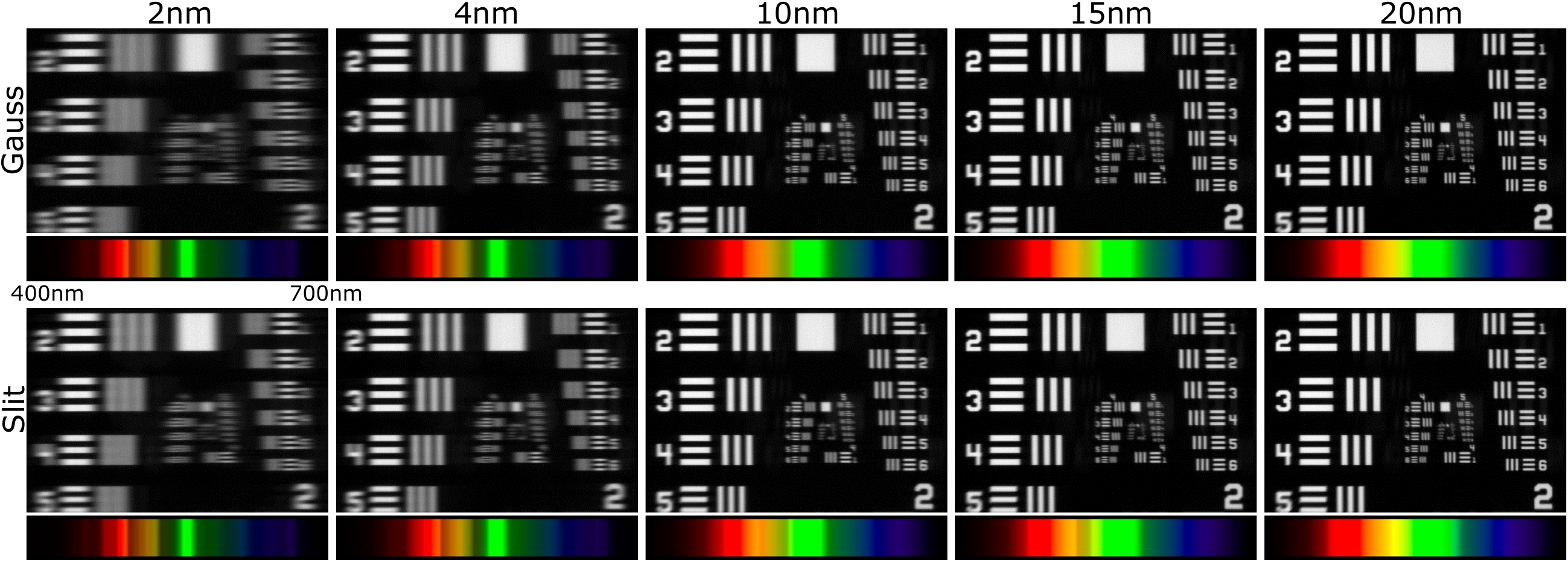}
	\caption{\small \textbf{Visualization of spectral and spatial resolutions.} We illuminated a USAF resolution target with a CFL lamp. We varied aperture types and widths to get spatial and spectral measurements. Each row shows image and spectrum for a specific aperture, and each column shows image and spectrum for a fixed spectral standard deviation. The results clearly illustrate the tradeoff between spatial and spectral resolution.}
	\label{fig:real_usaf}
\end{figure}
\paragraph{Optical setup.} We built an optical setup shown in Fig.\ \ref{fig:setup} with relevant components marked.
The setup is a minor modification of the schematic shown in Fig.\ \ref{fig:schematic}.
We placed a spatial light modulator (SLM) on plane P2 which enabled display of various coded apertures, and a diffraction grating in plane P3.
The spectral measurement camera is on plane P4.
Instead of placing spatial camera on P5, we place it on P3 (using beamsplitter BS1). Since we do not code the rainbow plane P4, image on P3 and P5 are equivalent.
Focal length of all our lenses was $75$ mm and the diffraction grating had $300$ grooves/mm.

\paragraph{Visualization of spectral and spatial resolutions.} To illustrate our hypothesis, we placed a USAF resolution chart on the image plane P1.
The scene was illuminated with a cool white compact fluorescent lamp (CFL) which is comprised of several narrow peaks.
This setup enabled us to simultaneously visualize sharp spectrum as well as sharp spatial features.
Figure \ref{fig:real_usaf} shows images and spectra for some representative cases. Each row shows results for a specific coded aperture, whereas each column shows results for a fixed spectral resolution.
The trend of decreasing spectral resolution with increasing spatial resolution is clearly visible.
Further, a Gaussian aperture is superior to slit in terms of greater spatial resolution for the same spectral resolution, which agrees with our theoretical findings.

\begin{figure}[!tt]
	\centering
	\includegraphics[width=0.5\columnwidth]{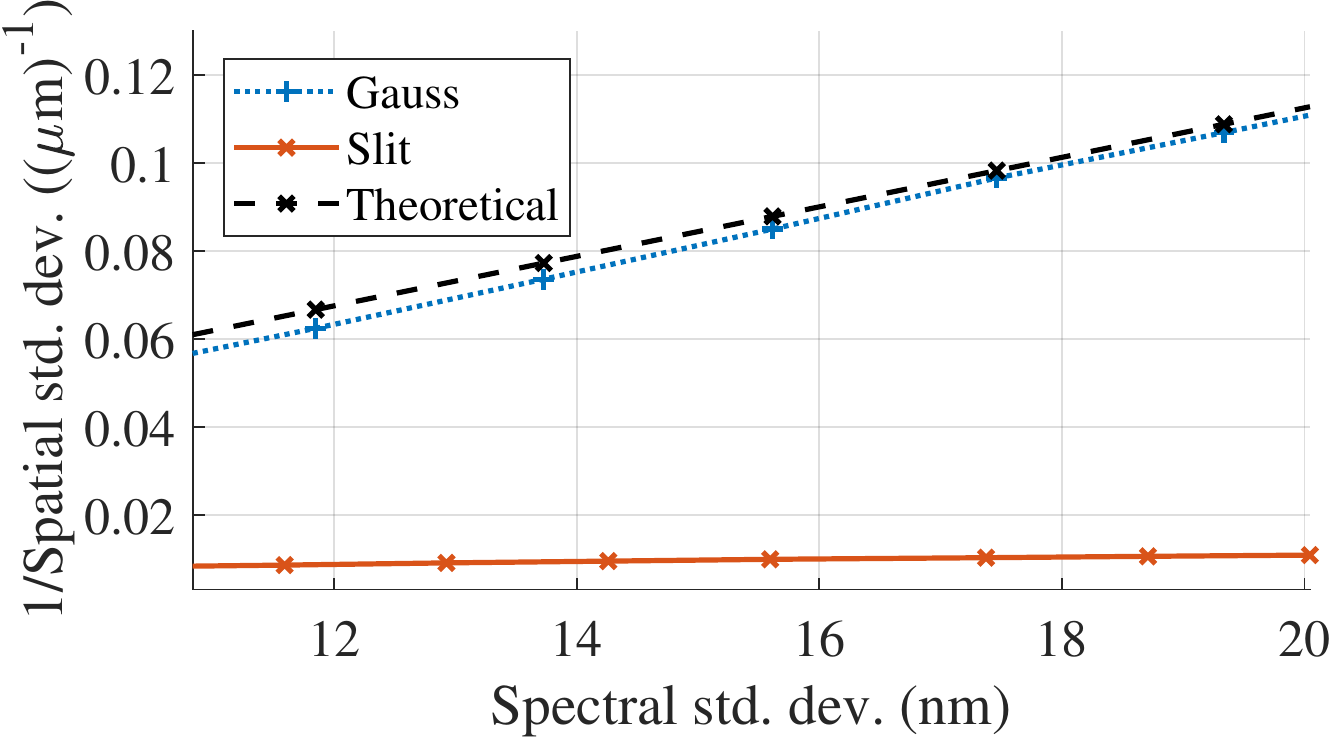}
	\caption{\small \textbf{Quantitative measurement of resolutions.} We captured spatial and spectral measurements using the setup in Fig. \ref{fig:setup}. We illuminated a pinhole with a narrowband light source at 670 nm and swept across various aperture shapes and sizes. We then plotted reciprocal of spatial standard deviation against spectral standard deviation, clearly showing a straight line.}
	\label{fig:real_fwhm}
\end{figure}

\begin{figure}[!tt]
	\centering
	\begin{subfigure}[c]{\columnwidth}
		\centering
		\includegraphics[width=\columnwidth]{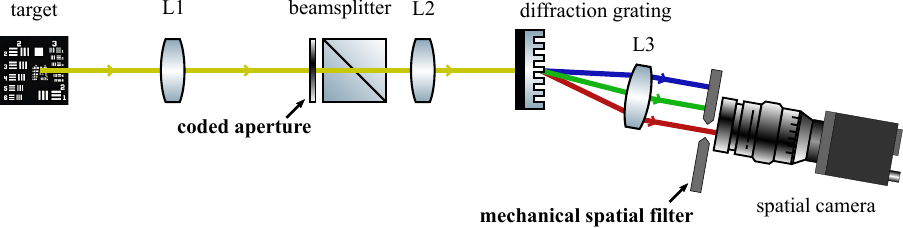}
	\end{subfigure}
	\\
	\begin{subfigure}[c]{\columnwidth}
		\centering
		\includegraphics[width=\columnwidth]{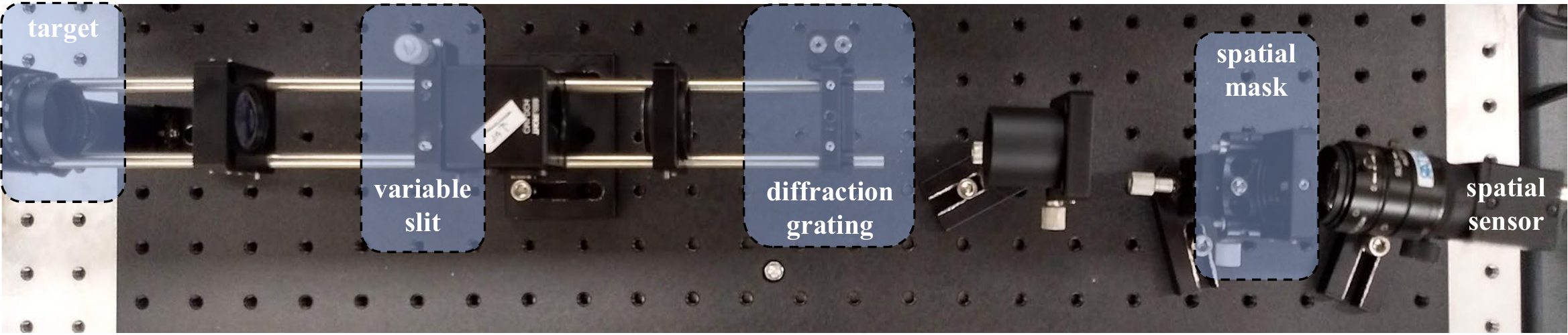}
	\end{subfigure}
	\caption{\small \textbf{Schematic and image of our prototype for spectral programming}. We illuminated a USAF target with 520 nm and 532 nm lasers. We then blocked 532 nm with a spatial mask on the rainbow plane. Images were captured with various slit widths to show the effect of space-spectrum uncertainty.}
	\label{fig:prog_setup}
\end{figure}

\paragraph{Quantitative verification.}
We illuminated a pinhole with a spectrally-narrowband light source with a central wavelength of 670 nm and an FWHM of 3 nm. 
We then captured both spectrum of the light source and image of the pinhole, which we then used for computing the corresponding standard deviations.
%
%
Figure \ref{fig:real_fwhm} compares reciprocal of spatial resolution against spectral resolution. The two plots show a straight line, thereby verifying that the product of spatial and spectral resolutions is a constant.
We also observe that the line for Gaussian aperture is very close to the theoretically optimal line, thereby confirming that the lower bound is tight, even in practice.

\subsection{Spectral Programming}
\begin{figure}[!tt]
	\centering
	\begin{subfigure}[t]{0.05\columnwidth}
		\centering
		\rotatebox{90}{\small Illum.}
	\end{subfigure}
	\begin{subfigure}[t]{0.18\columnwidth}
		\centering
		\includegraphics[width=\columnwidth]{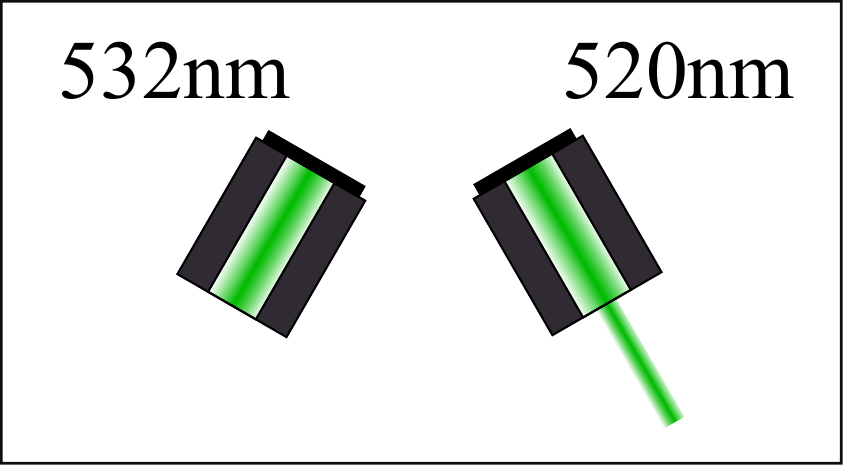}
	\end{subfigure}
	\begin{subfigure}[t]{0.18\columnwidth}
		\centering
		\includegraphics[width=\columnwidth]{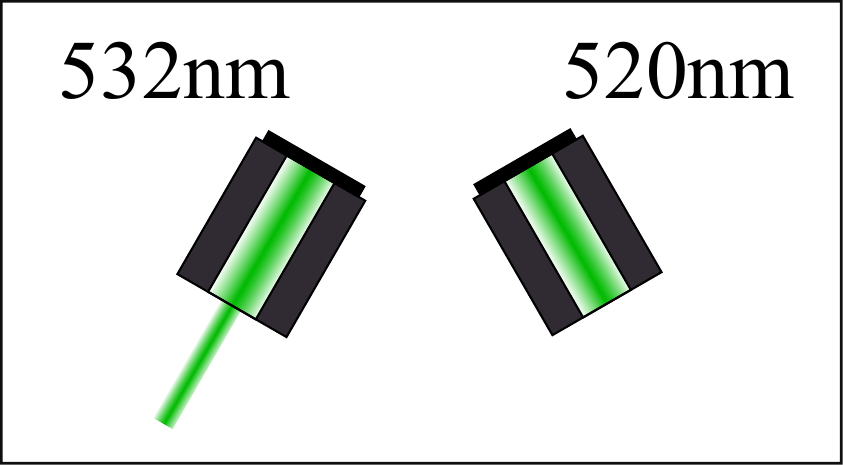}
	\end{subfigure}
	\begin{subfigure}[t]{0.18\columnwidth}
		\centering
		\includegraphics[width=\columnwidth]{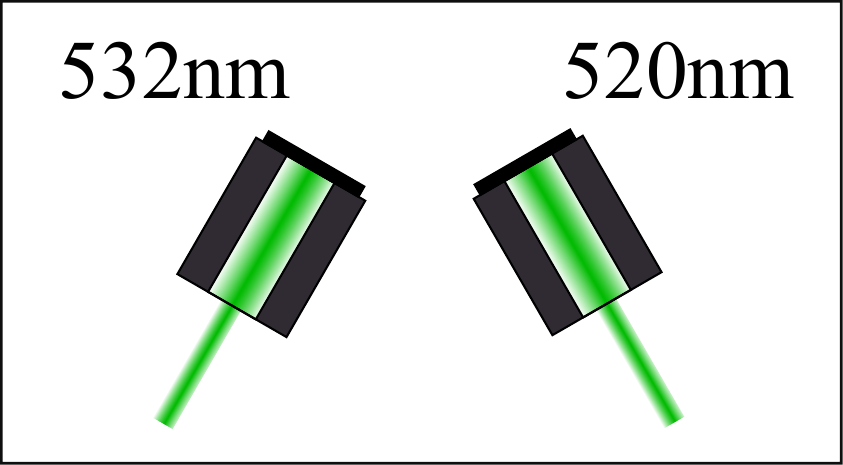}
	\end{subfigure}
	\begin{subfigure}[t]{0.18\columnwidth}
		\centering
		\includegraphics[width=\columnwidth]{figures/lasers_both.pdf}
	\end{subfigure}
	\begin{subfigure}[t]{0.18\columnwidth}
		\centering
		\includegraphics[width=\columnwidth]{figures/lasers_both.pdf}
	\end{subfigure}
	\\
	\begin{subfigure}[t]{0.05\columnwidth}
		\centering
		\rotatebox{90}{\small \qquad Image}
	\end{subfigure}
	\begin{subfigure}[t]{0.18\columnwidth}
		\centering
		\includegraphics[width=\columnwidth]{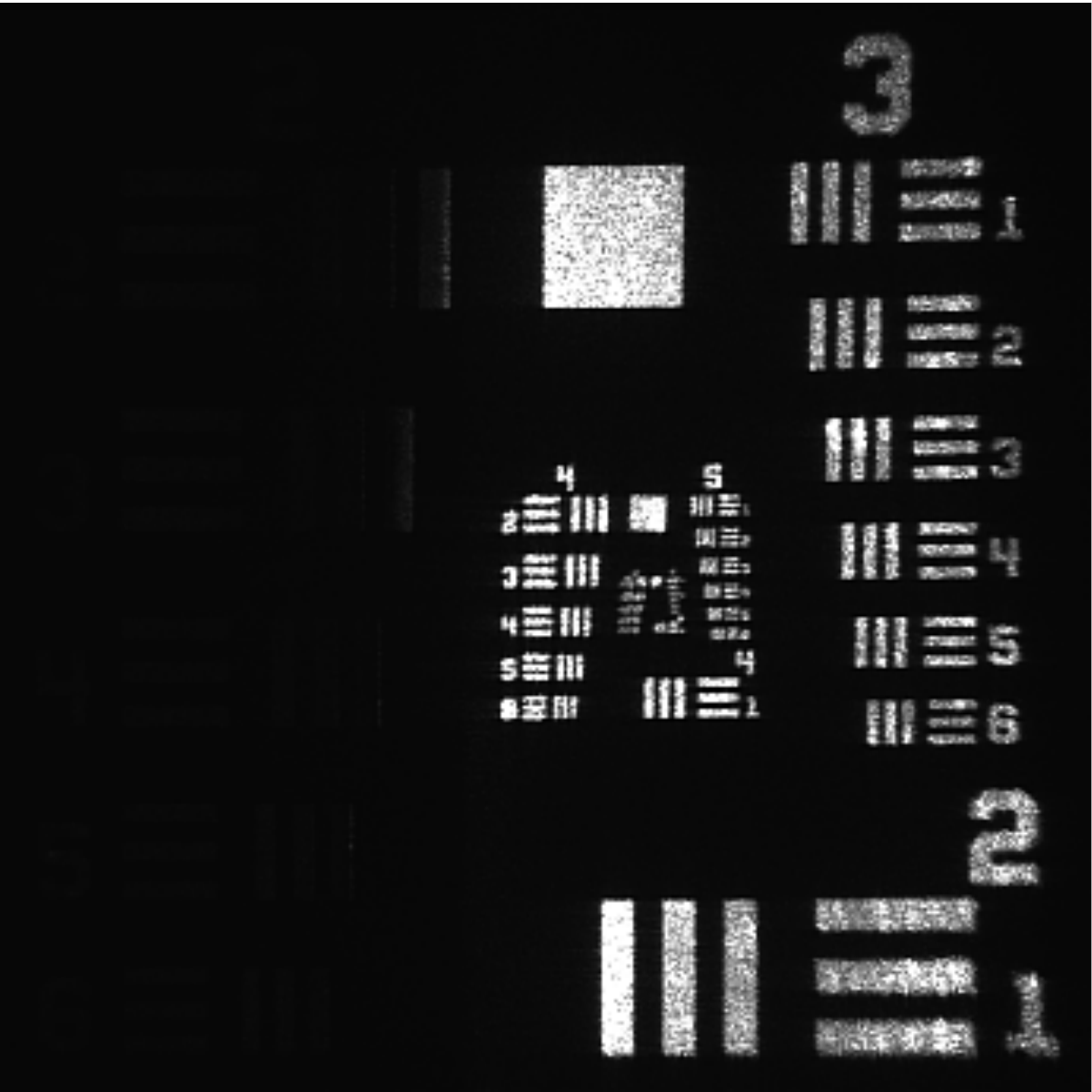}
	\end{subfigure}
	\begin{subfigure}[t]{0.18\columnwidth}
		\centering
		\includegraphics[width=\columnwidth]{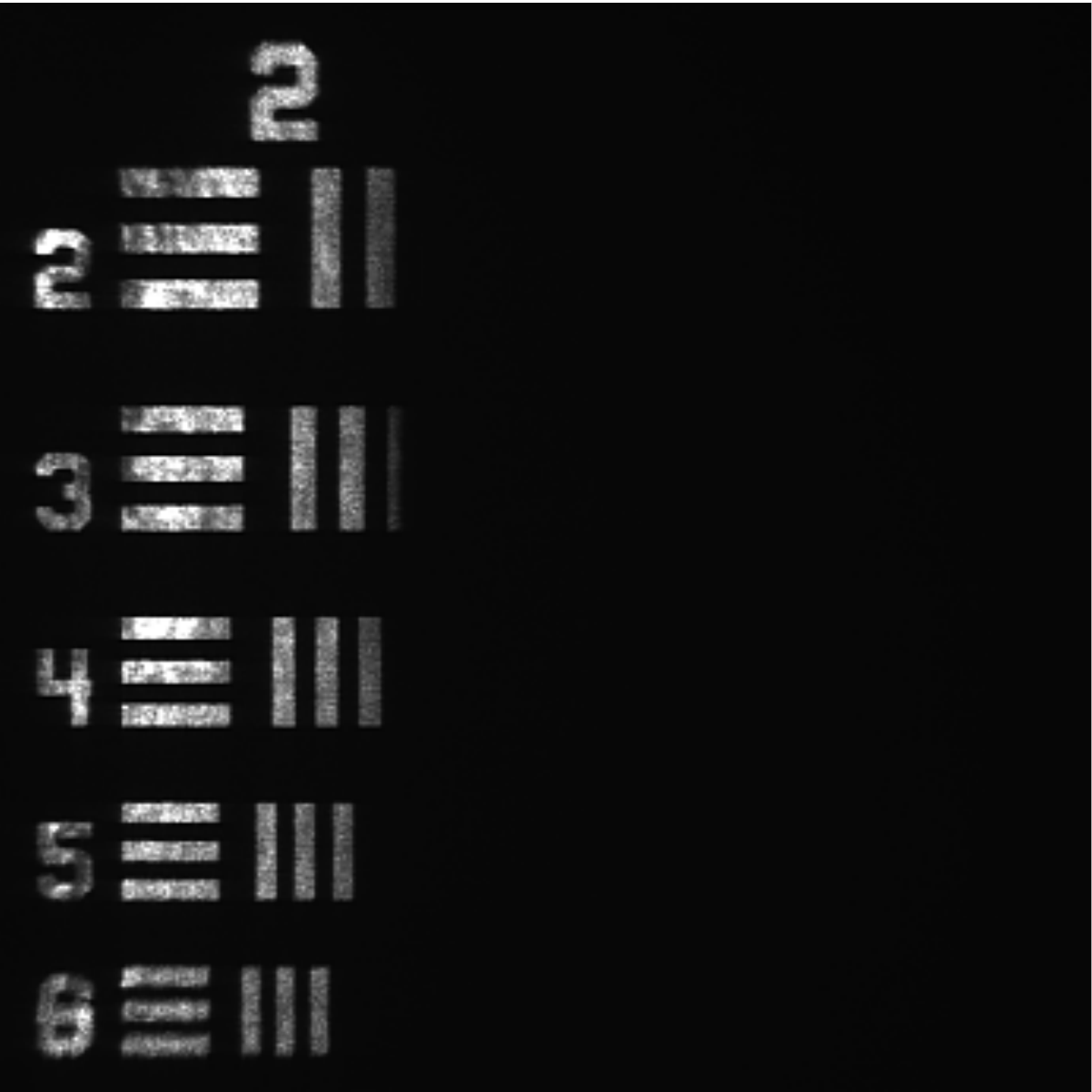}
	\end{subfigure}	
	\begin{subfigure}[t]{0.18\columnwidth}
		\centering
		\includegraphics[width=\columnwidth]{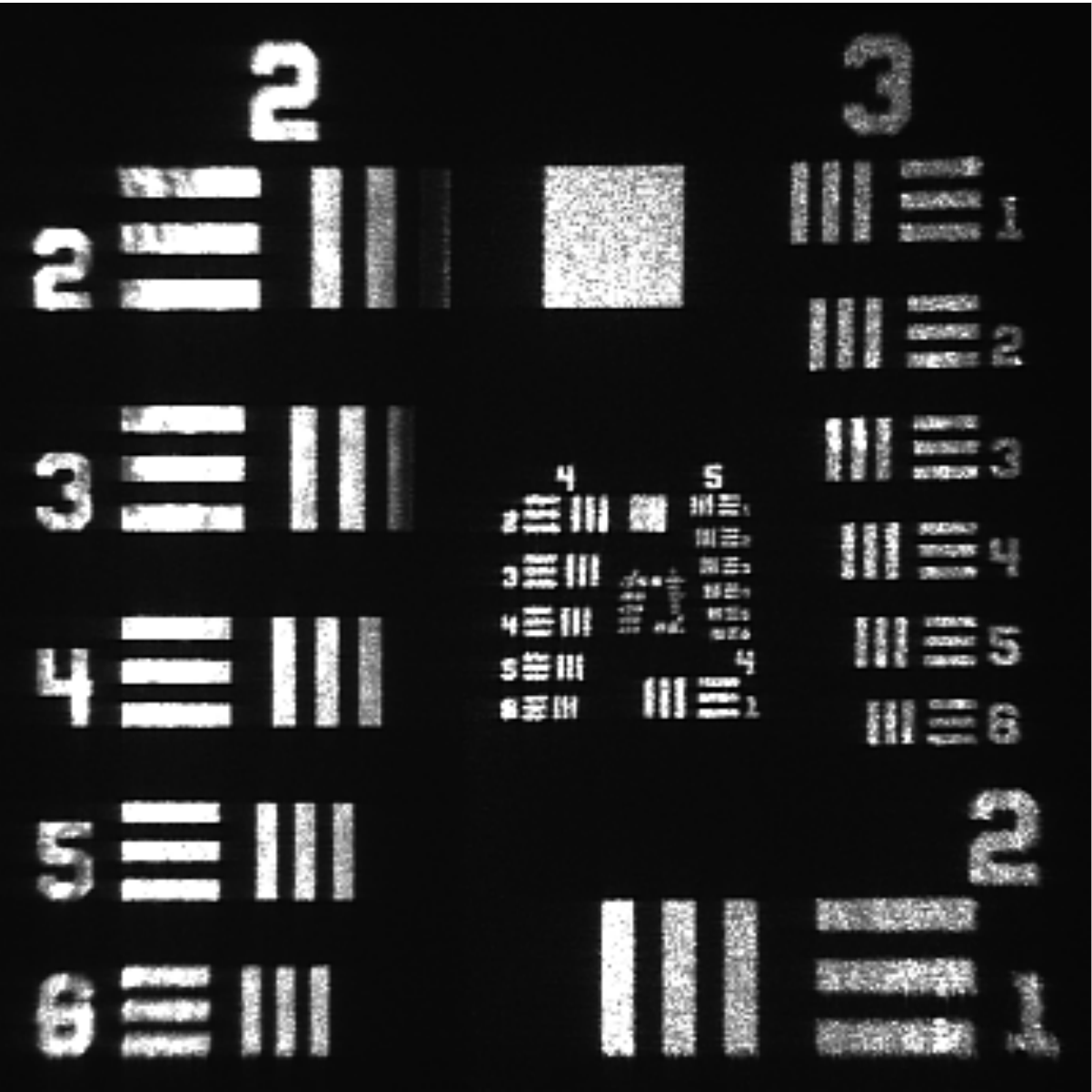}
	\end{subfigure}
	\begin{subfigure}[t]{0.18\columnwidth}
		\centering
		\includegraphics[width=\columnwidth]{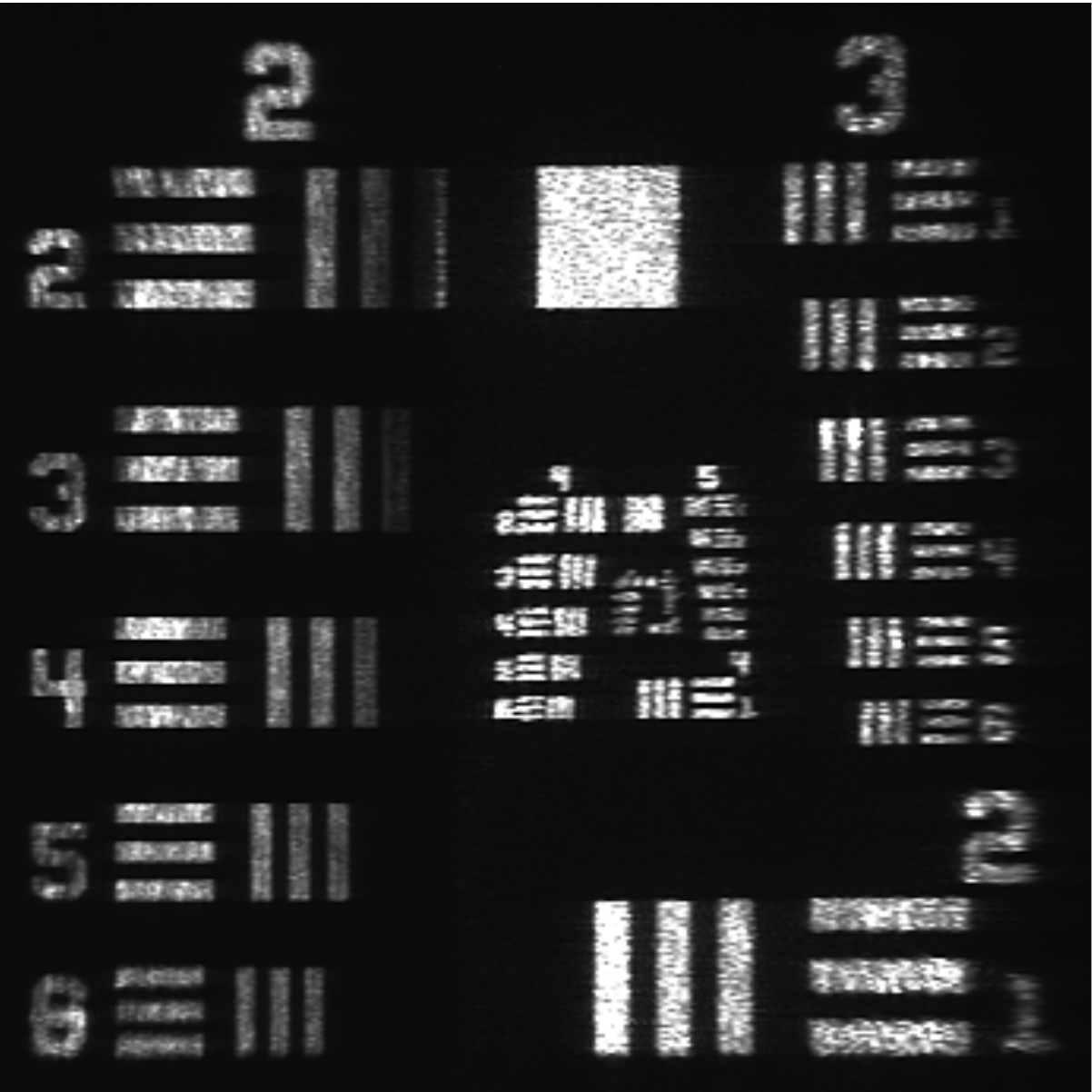}
	\end{subfigure}
	\begin{subfigure}[t]{0.18\columnwidth}
		\centering
		\includegraphics[width=\columnwidth]{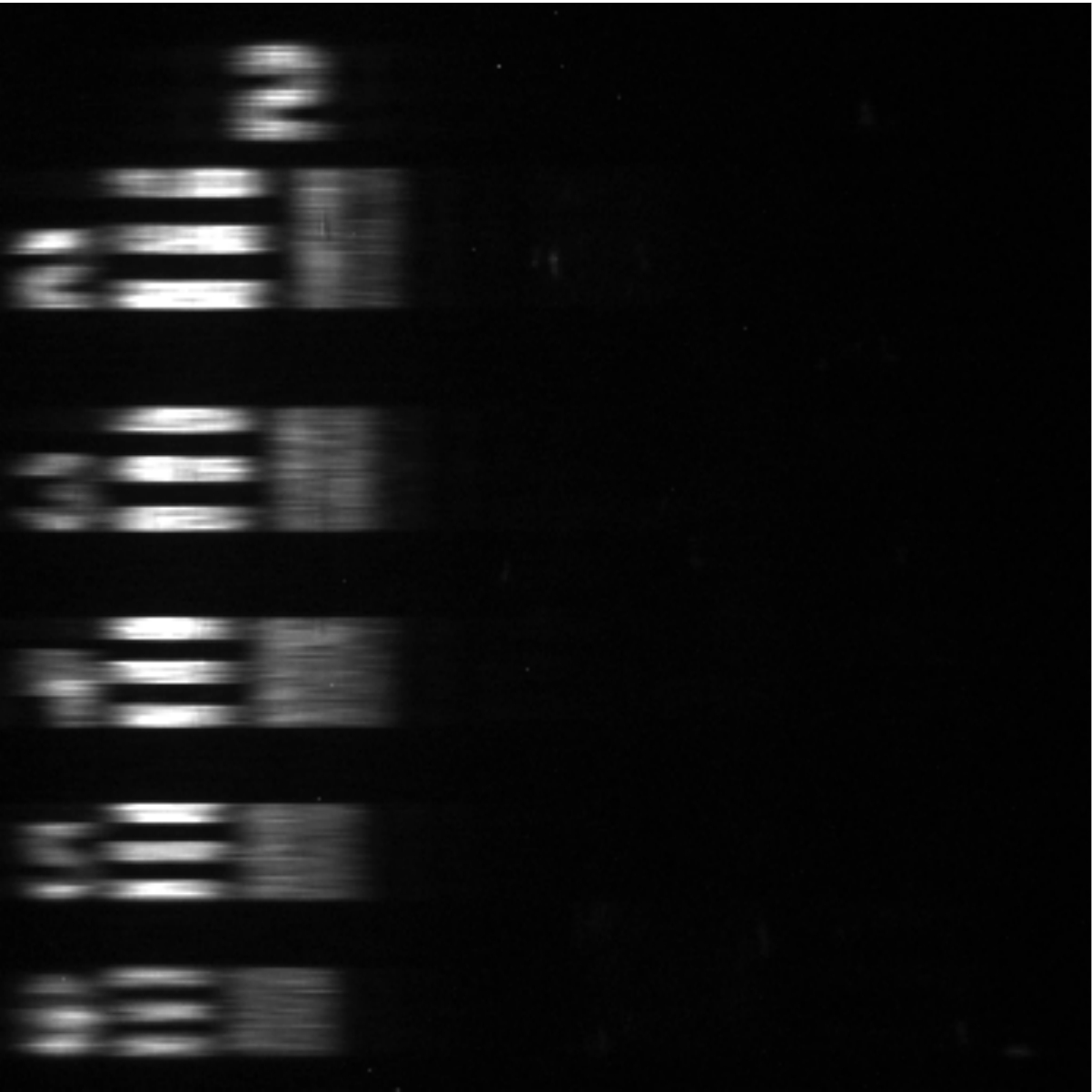}
	\end{subfigure}
	\\
	\begin{subfigure}[t]{0.05\columnwidth}
		\centering
		\rotatebox{90}{\small Aperture}
	\end{subfigure}
	\begin{subfigure}[t]{0.18\columnwidth}
		\centering
		\includegraphics[width=\columnwidth]{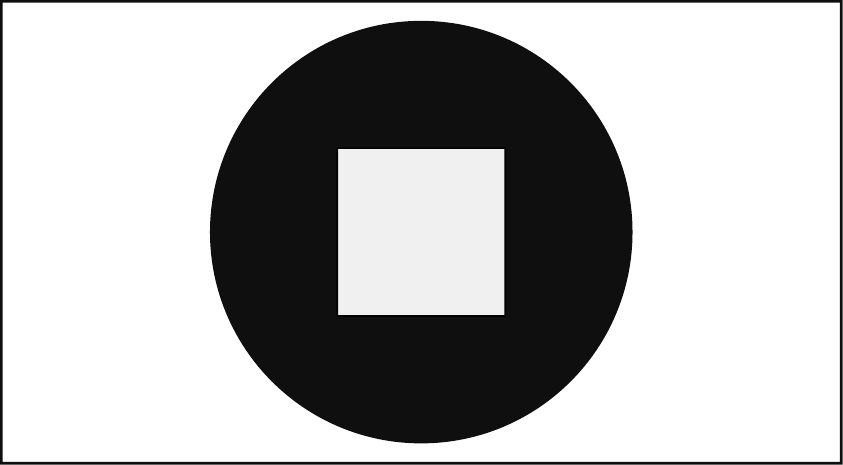}
	\end{subfigure}
	\begin{subfigure}[t]{0.18\columnwidth}
		\centering
		\includegraphics[width=\columnwidth]{figures/lasers_open.pdf}
	\end{subfigure}
	\begin{subfigure}[t]{0.18\columnwidth}
		\centering
		\includegraphics[width=\columnwidth]{figures/lasers_open.pdf}
	\end{subfigure}
	\begin{subfigure}[t]{0.18\columnwidth}
		\centering
		\includegraphics[width=\columnwidth]{figures/lasers_open.pdf}
	\end{subfigure}
	\begin{subfigure}[t]{0.18\columnwidth}
		\centering
		\includegraphics[width=\columnwidth]{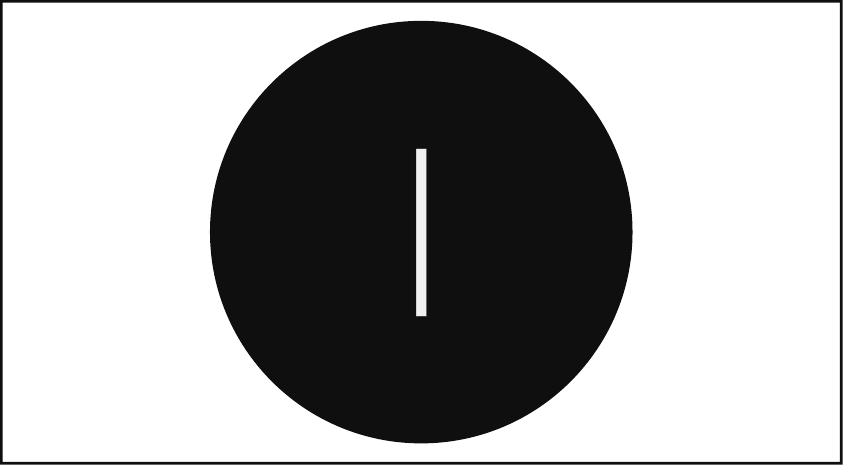}
	\end{subfigure}
	\\
	\begin{subfigure}[t]{0.05\columnwidth}
		\centering
		\rotatebox{90}{\small \quad Filter}
	\end{subfigure}
	\begin{subfigure}[t]{0.18\columnwidth}
		\centering
		\includegraphics[width=\columnwidth]{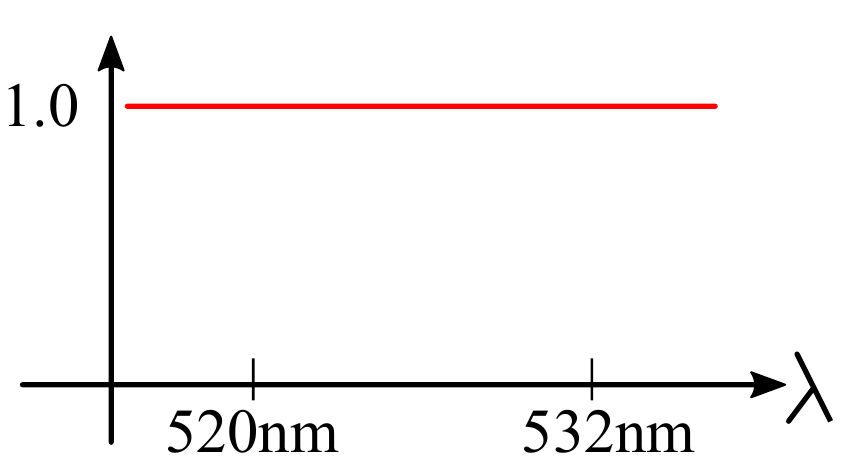}
	\end{subfigure}
	\begin{subfigure}[t]{0.18\columnwidth}
		\centering
		\includegraphics[width=\columnwidth]{figures/lasers_broadband.pdf}
	\end{subfigure}
	\begin{subfigure}[t]{0.18\columnwidth}
		\centering
		\includegraphics[width=\columnwidth]{figures/lasers_broadband.pdf}
	\end{subfigure}
	\begin{subfigure}[t]{0.18\columnwidth}
		\centering
		\includegraphics[width=\columnwidth]{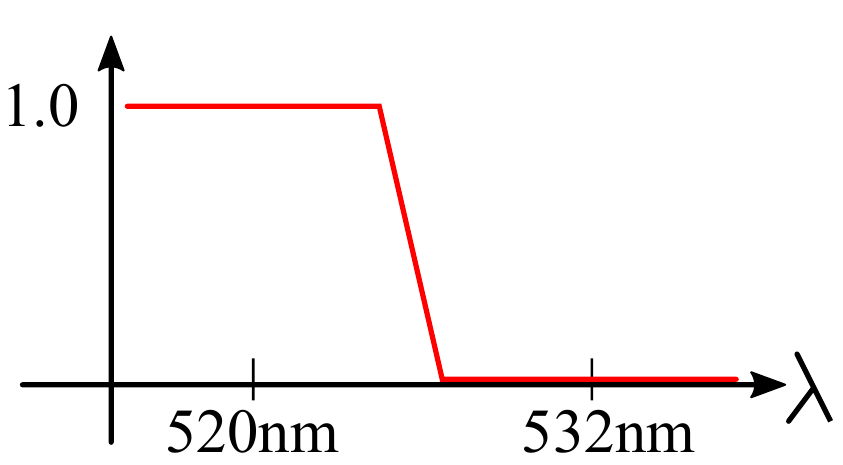}
	\end{subfigure}
	\begin{subfigure}[t]{0.18\columnwidth}
		\centering
		\includegraphics[width=\columnwidth]{figures/lasers_cutoff.pdf}
	\end{subfigure}
	\\
	\begin{subfigure}[t]{0.05\columnwidth}
		\centering
		\rotatebox{90}{\small \qquad Spectrum}
	\end{subfigure}
	\begin{subfigure}[t]{0.18\columnwidth}
		\centering
		\includegraphics[width=\columnwidth]{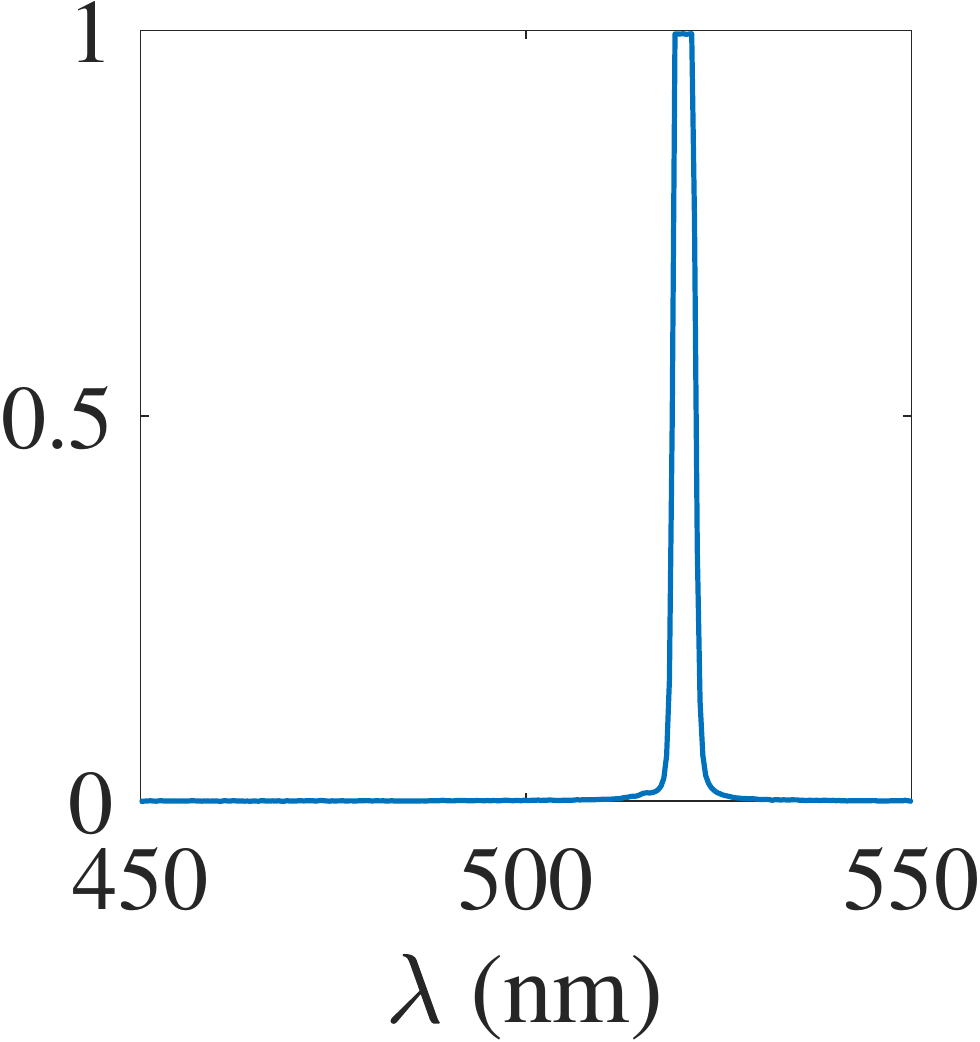}
	\end{subfigure}
	\begin{subfigure}[t]{0.18\columnwidth}
		\centering
		\includegraphics[width=\columnwidth]{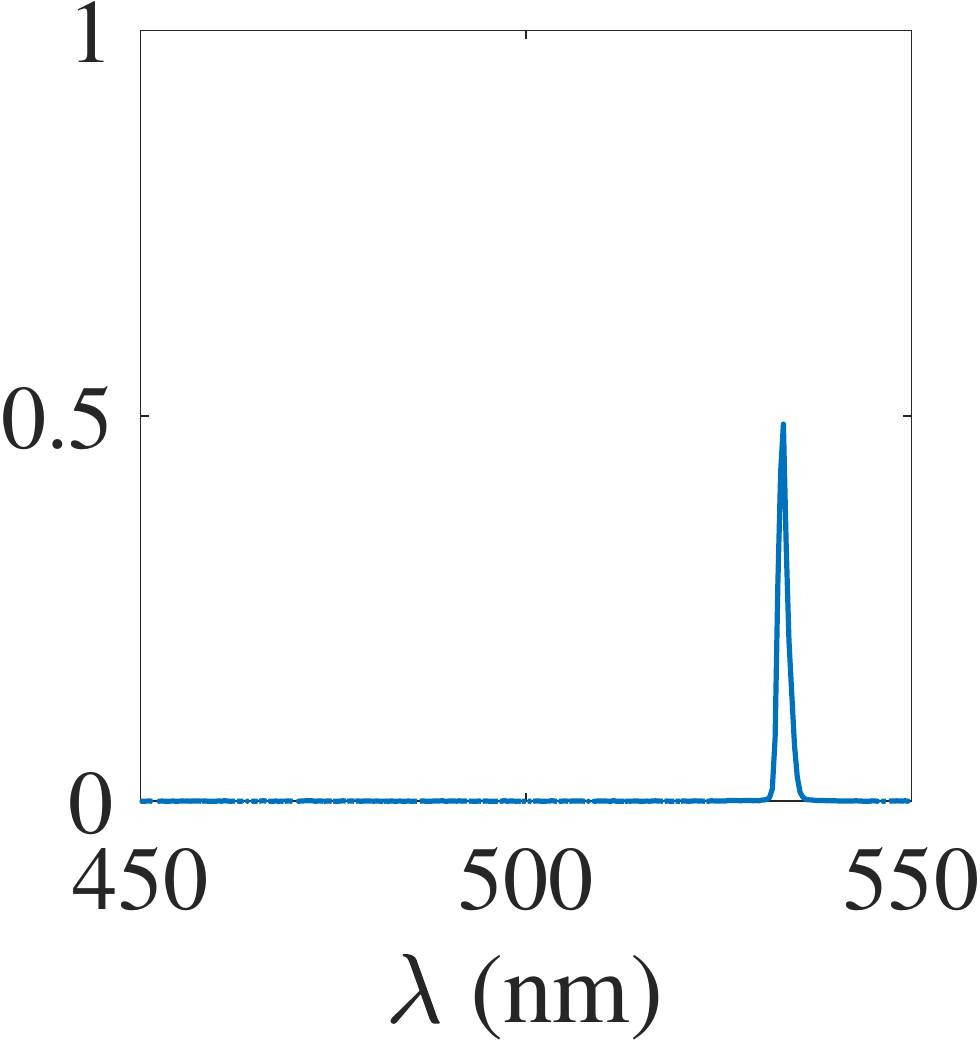}
	\end{subfigure}	
	\begin{subfigure}[t]{0.18\columnwidth}
		\centering
		\includegraphics[width=\columnwidth]{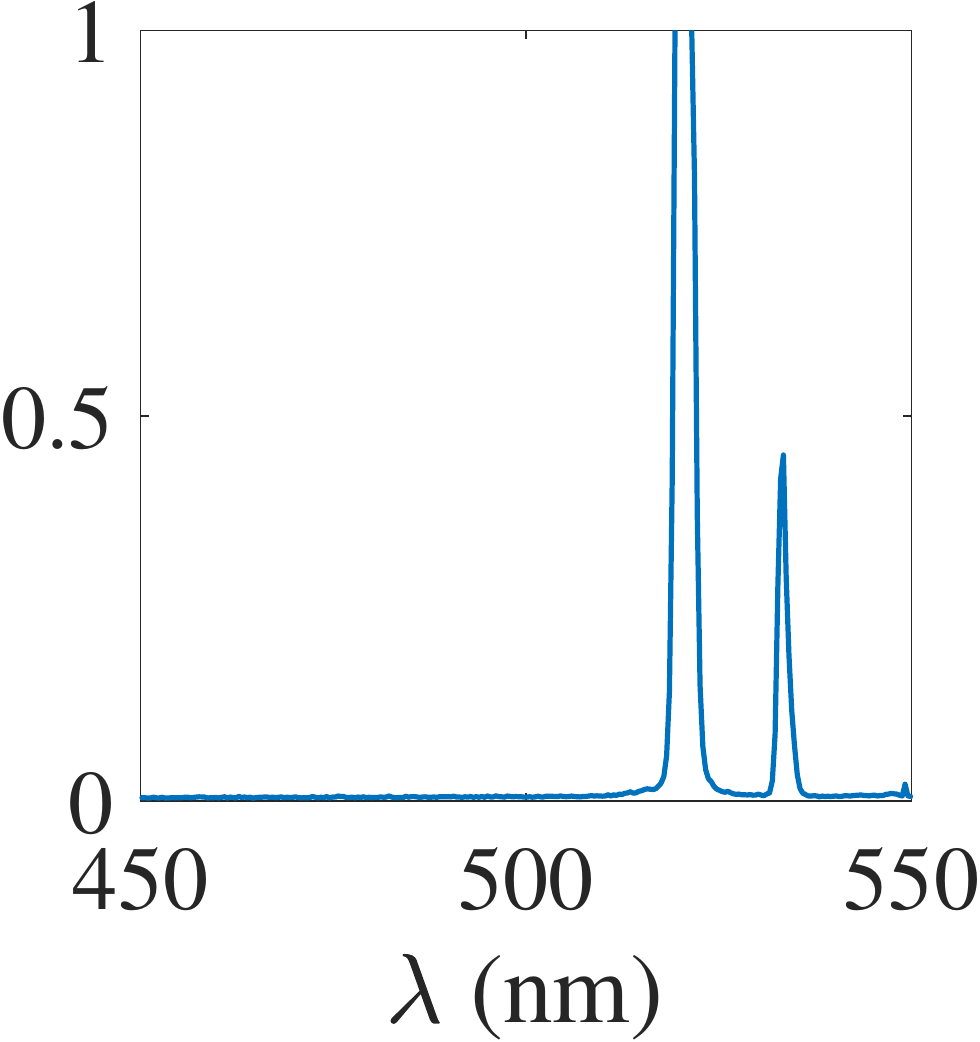}
	\end{subfigure}
	\begin{subfigure}[t]{0.18\columnwidth}
		\centering
		\includegraphics[width=\columnwidth]{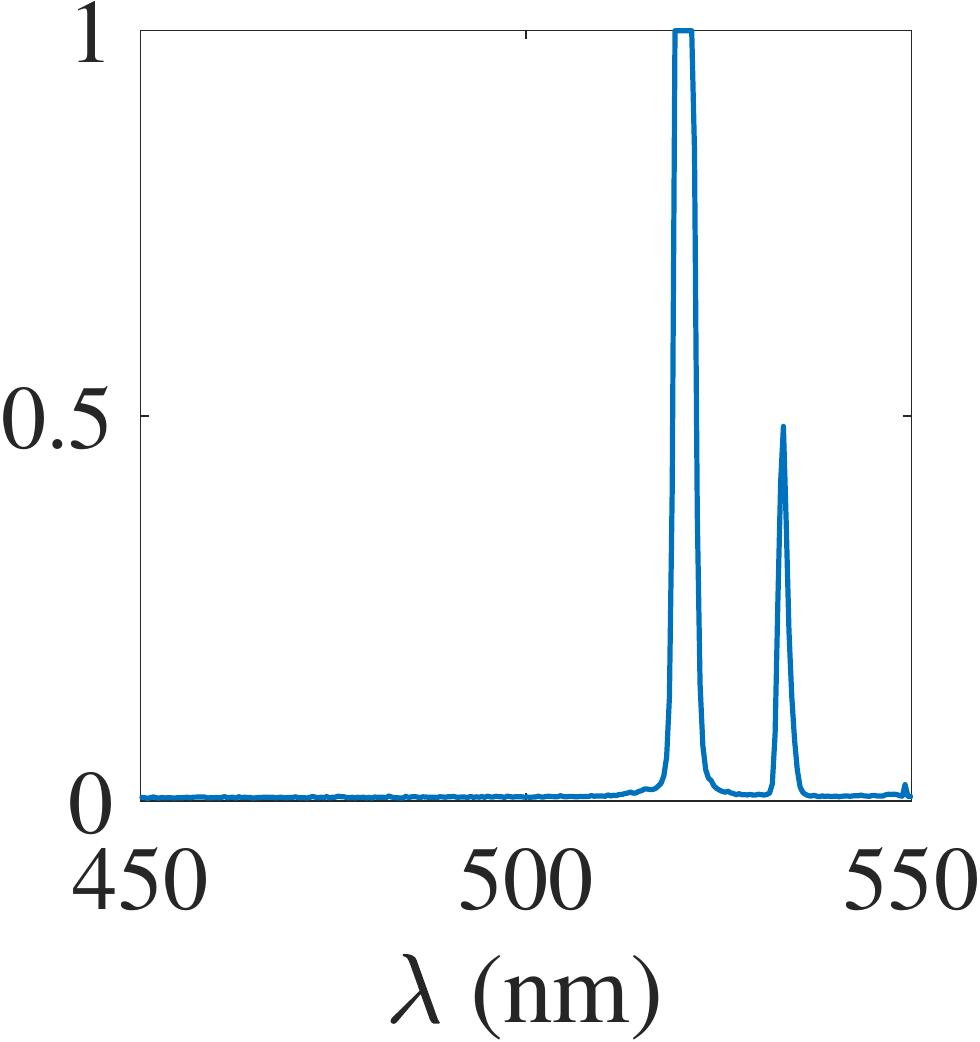}
	\end{subfigure}
	\begin{subfigure}[t]{0.18\columnwidth}
		\centering
		\includegraphics[width=\columnwidth]{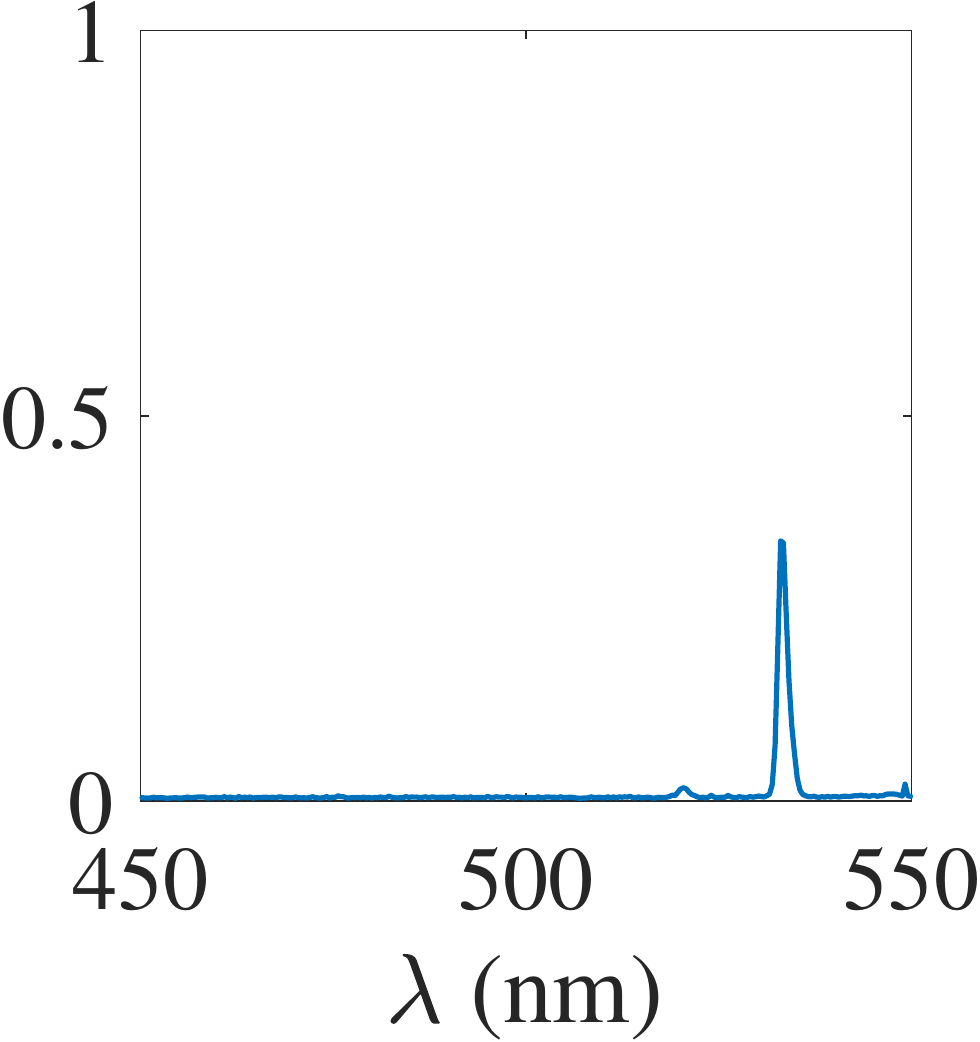}
	\end{subfigure}
	\caption{\small \textbf{Spectral filtering with narrowband sources.} A consequence of space-spectrum bandwidth product is the incapability of spectral programming at high resolution. In this example, we show how blocking one of the two closely spaced narrowband lasers and only be done with severe loss in resolution.}
	\label{fig:narrowband_programming}
\end{figure}

\paragraph{Effect of edge-pass filter.} The tradeoff between spectral and spatial resolution affects how well the spectrum can be coded.
To test this, we illuminate two closely spaced spatial points in a scene with two narrowband light sources (520 nm and 532 nm).
We then attempt to block the 520 nm laser with various coded apertures and observe the spatial image.
Figure \ref{fig:prog_setup} shows the schematic and our lab prototype for spectral programming.
The setup is built very similar to the schematic shown in Fig.\ \ref{fig:schematic} with a spatial mask placed in plane P4.
The results are shown in Fig.\ \ref{fig:narrowband_programming}. With a broad aperture, it is not possible to effectively block one of the two lasers, shown in fourth column. 
A narrow aperture can lead to effective blocking (compare first and last columns) but with loss in resolution.

\begin{figure}[!tt]
	\centering
	\begin{subfigure}[c]{0.48\textwidth}
		\centering
		\begin{subfigure}[c]{0.48\textwidth}
			\centering
			\includegraphics[width=\textwidth]{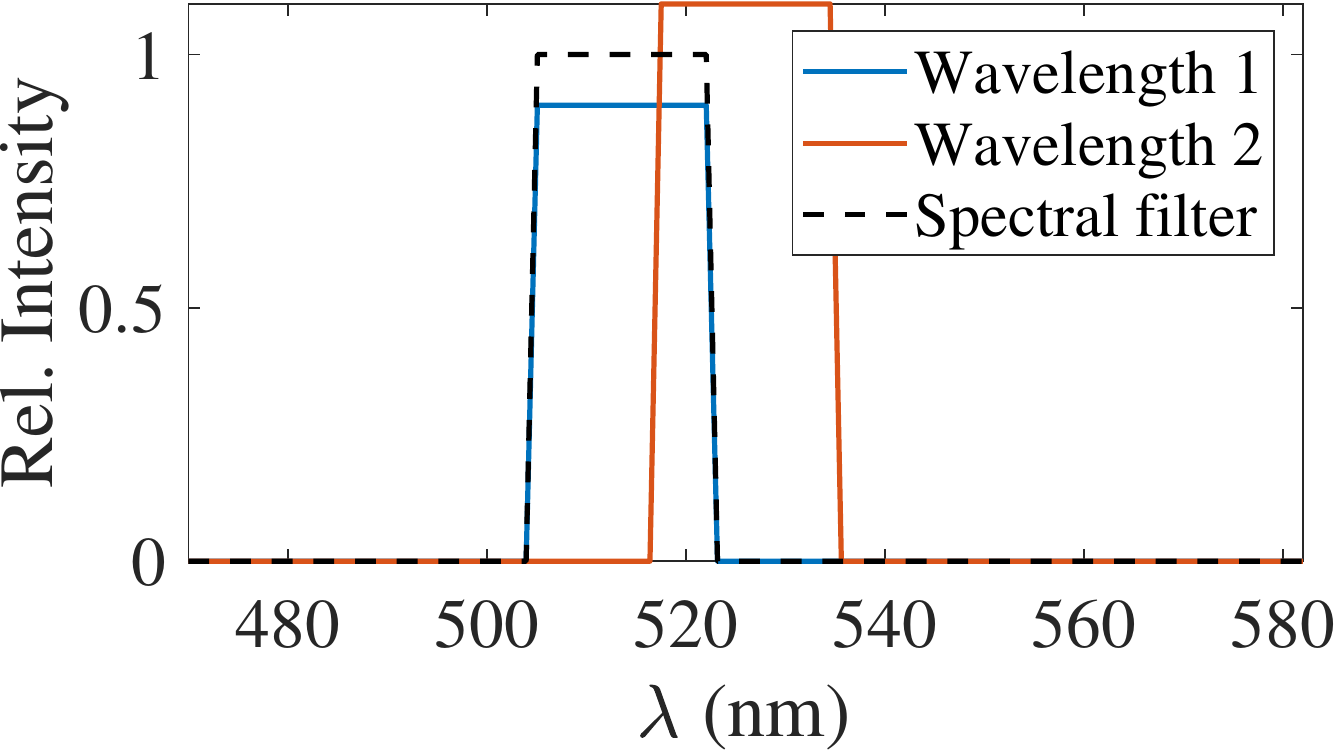}
		\end{subfigure}
		\begin{subfigure}[c]{0.48\textwidth}
			\centering
			\includegraphics[width=\textwidth]{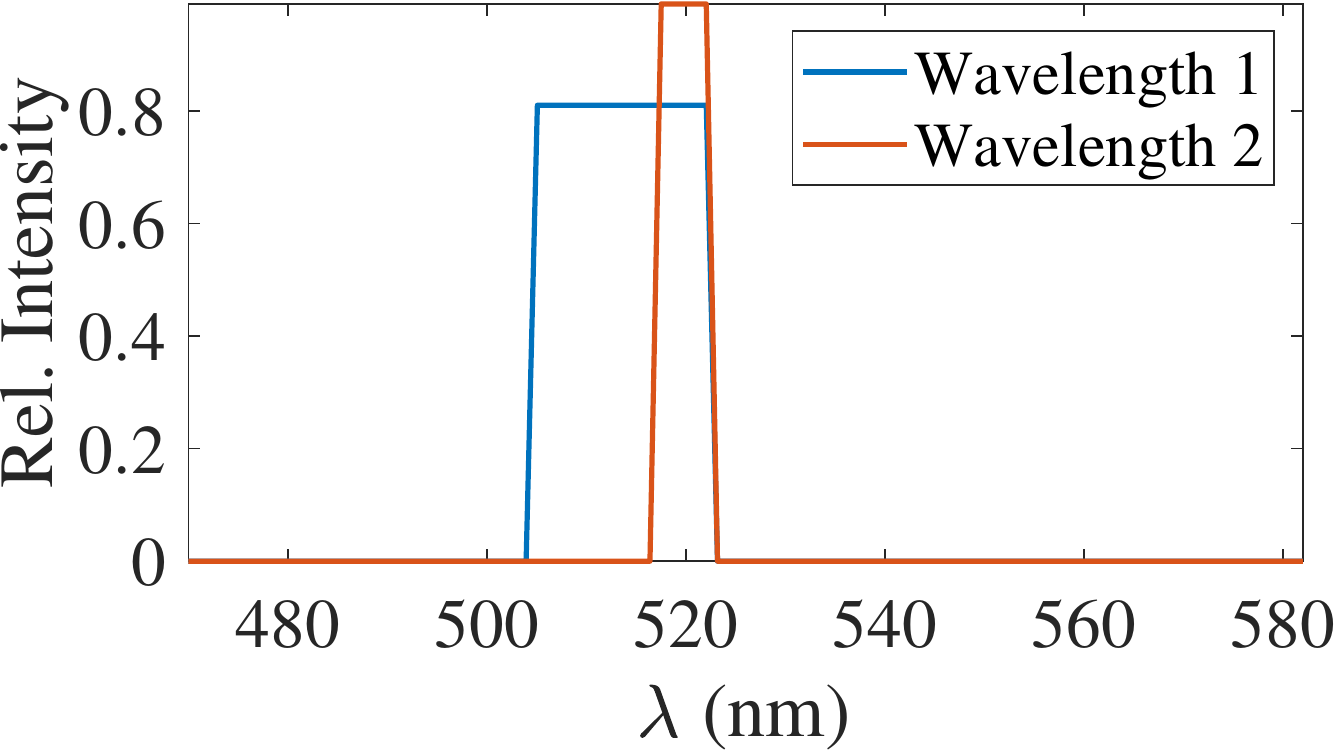}
		\end{subfigure}
		\caption{\small Slit; spectrum before and after spectral filtering}
	\end{subfigure}
	\begin{subfigure}[c]{0.48\textwidth}
		\centering
		\begin{subfigure}[c]{0.48\textwidth}
			\centering
			\includegraphics[width=\textwidth]{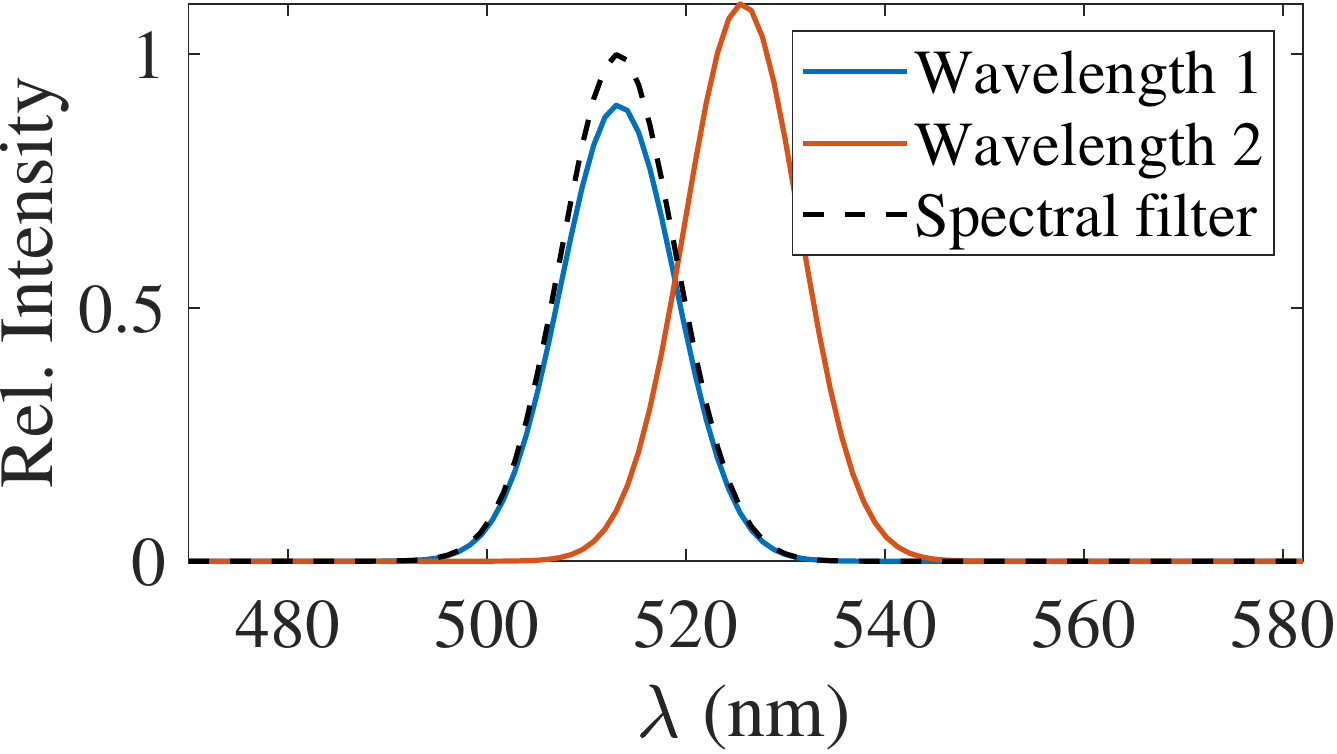}
		\end{subfigure}
		\begin{subfigure}[c]{0.48\textwidth}
			\centering
			\includegraphics[width=\textwidth]{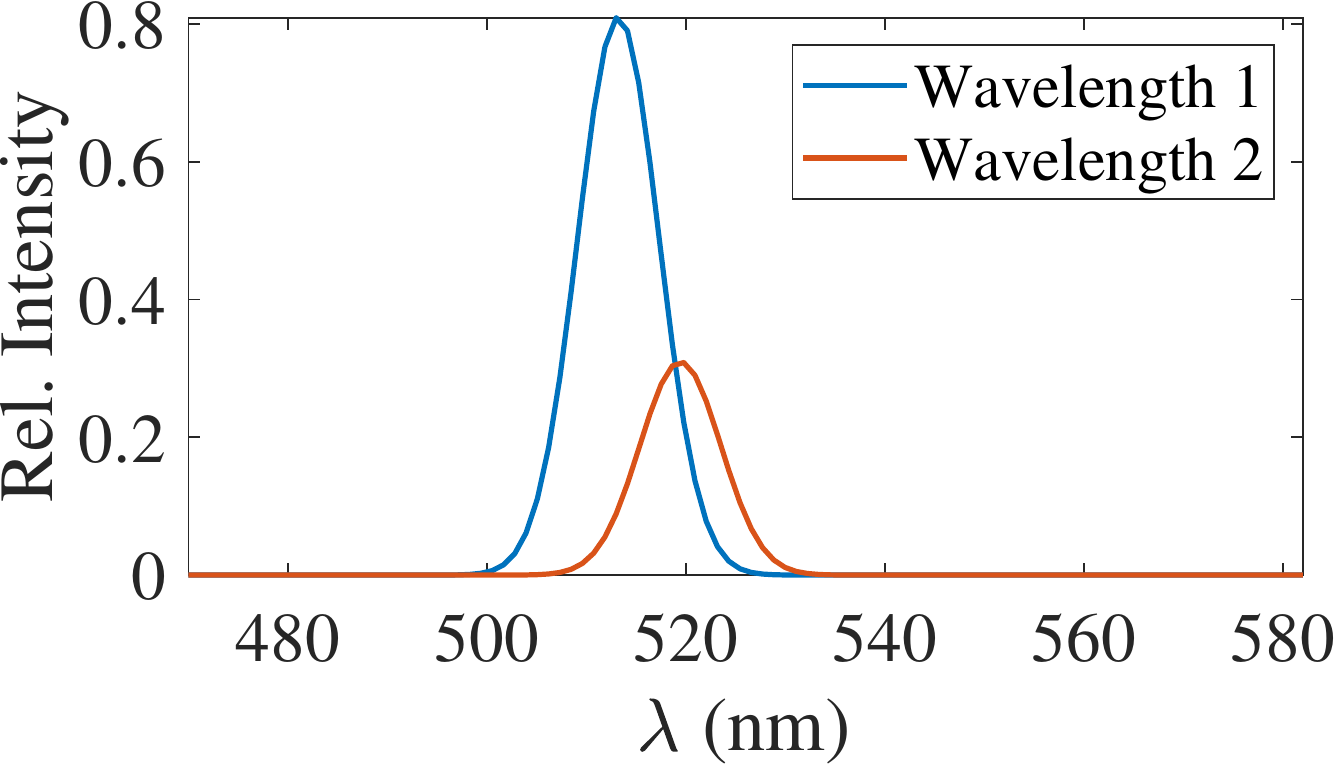}
		\end{subfigure}
		\caption{\small Gaussian; spectrum before and after filtering}
	\end{subfigure}
	\begin{subfigure}[c]{0.48\columnwidth}
		\centering
		\includegraphics[width=\columnwidth]{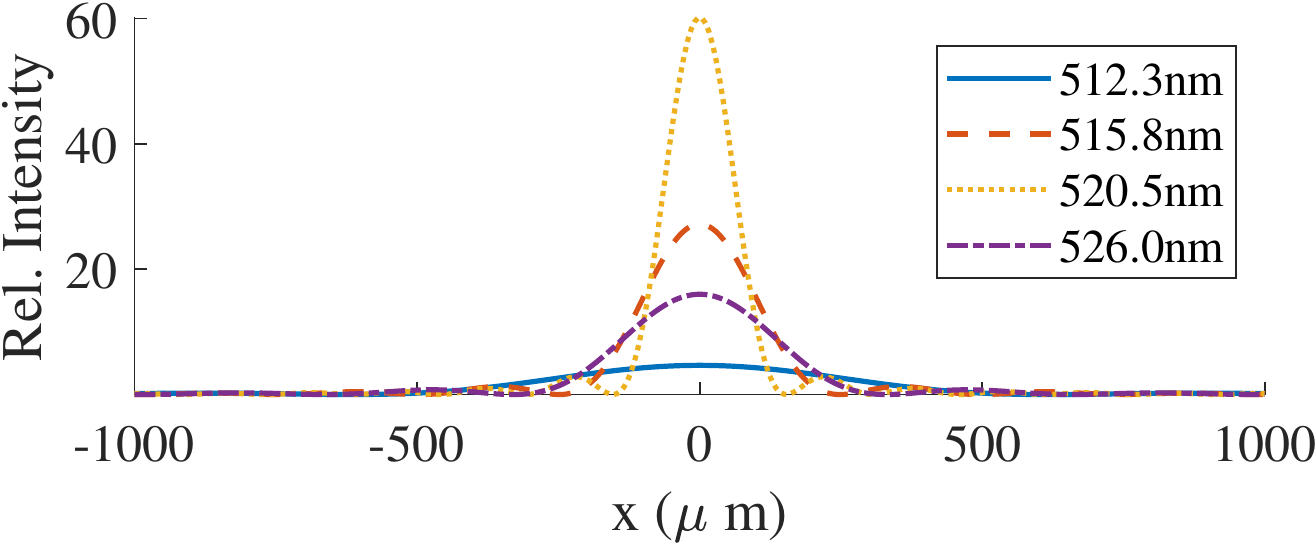}
		\caption{Slit + Slit}
	\end{subfigure}
	\begin{subfigure}[c]{0.48\columnwidth}
		\centering
		\includegraphics[width=\columnwidth]{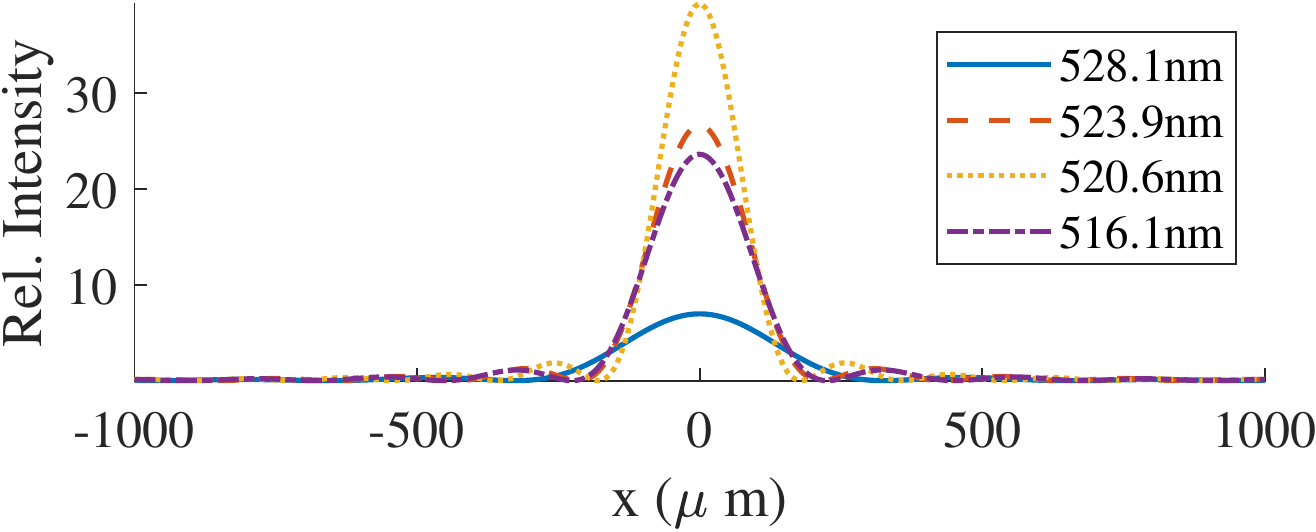}
		\caption{Gaussian + Slit}
	\end{subfigure}
	\begin{subfigure}[c]{0.48\columnwidth}
		\centering
		\includegraphics[width=\columnwidth]{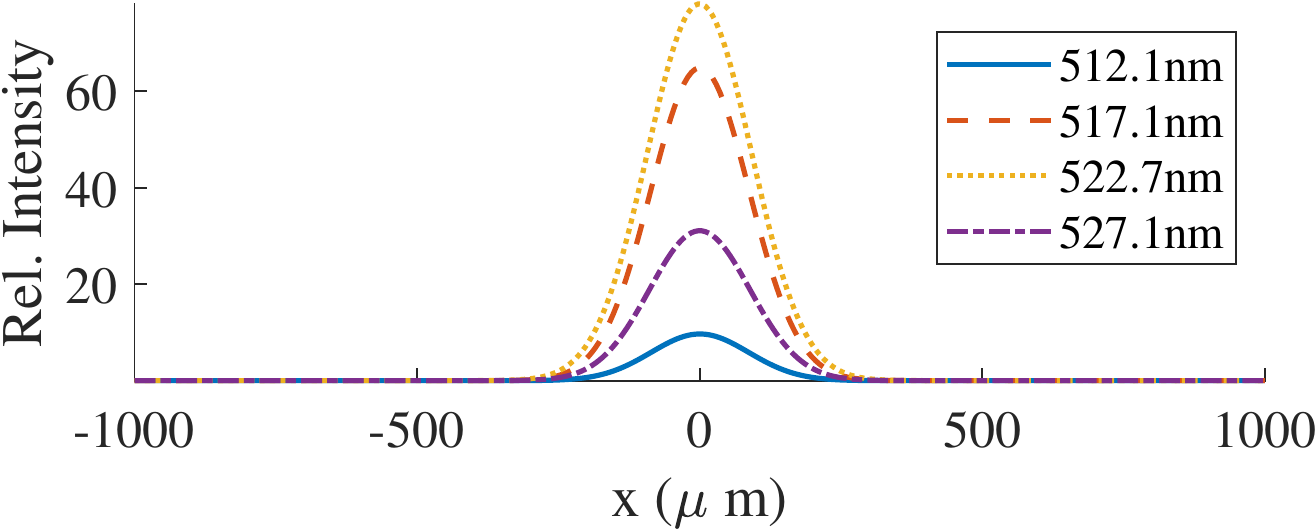}
		\caption{Gaussian + Gaussian}
	\end{subfigure}
	\begin{subfigure}[c]{0.48\columnwidth}
		\centering
		\includegraphics[width=\columnwidth]{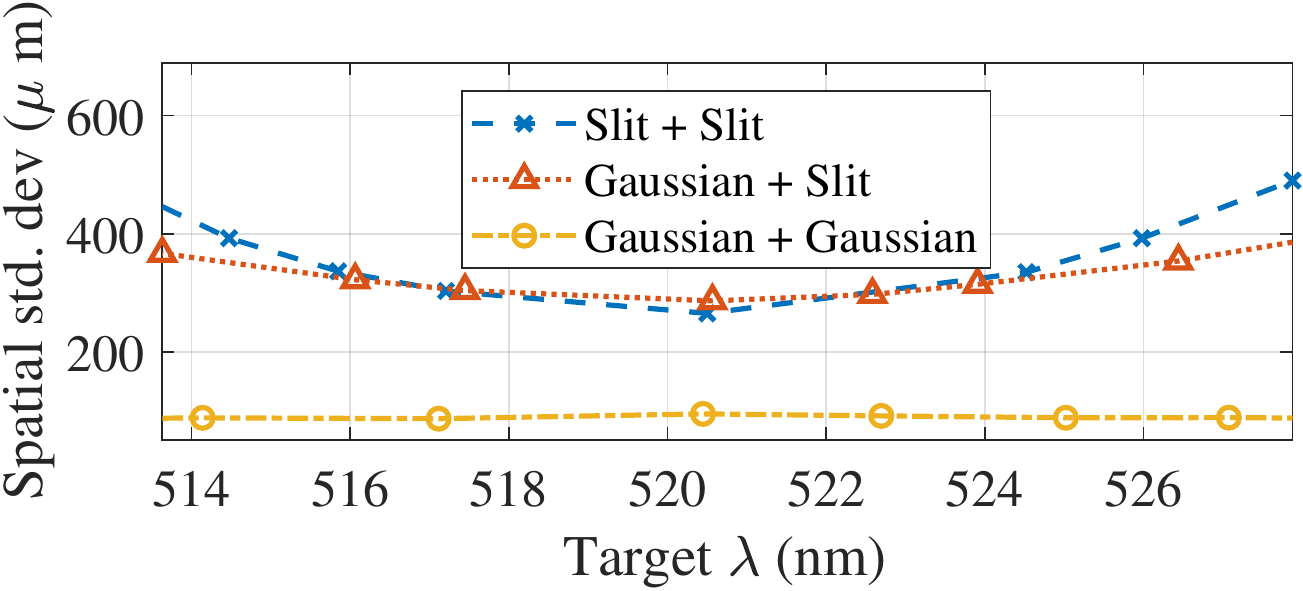}
		\caption{Resolution vs. wavelength}
	\end{subfigure}
	\caption{\small \textbf{Effect of narrow-band spectral filter on spatial resolution.} Spatial resolution is affected by the pupil code as well as the spatial mask used for creating a narrow-band spectral filter. We illuminated a pinhole with a 520 nm laser and then swept various spatial masks to filter wavelengths around 520 nm. For configurations where either the pupil code or the spatial mask was a slit, the spatial resolution got worse with increasing gap between desired and laser wavelength. In contrast, a configuration with both masks being Gaussian resulted in a wavelength-independent spatial blur.}
	\label{fig:filterpos}
\end{figure}
\paragraph{Effect of the shape of a narrowband filter.} We now show that a slit has unintended implication,  when used as a narrow band filter.
In order to perform narrowband spectral filtering, an intuitive choice is to place a narrow slit on the rainbow plane P4 and a camera on plane P5. 
This results in a spectra that looks similar to the example in Fig.\ \ref{fig:filterpos}(a). 
While such a  mask works well for the target wavelength, the spatial images corresponding to adjacent wavelengths have severe loss in spatial resolution.

To understand the effect of a narrowband filter, consider a scene illuminated by a monochromatic light source of wavelength $\lambda_1$. 
The resulting field on rainbow plane,
\begin{align}
i_4(x) = a\left(x - \lambda_1f\nu_0\right).
\end{align}
Now let a spatial mask $\widehat{a}(x)$ centered around $\lambda_2$ be placed on the rainbow plane.
Then the output just after the mask,
\begin{align}
\widetilde{i}_4(x) = a\left(x - \lambda_1f\nu_0\right)\widehat{a}\left(x - \lambda_2f\nu_0\right).
\end{align}
If both $a(x)$ and $\widehat{a}(x)$ are slits of width $W$, then the effective width is $W - |\lambda_1-\lambda_2|$, which decreases with increasing gap between the two wavelengths. 
This is illustrated by the plots of PSFs in Fig.\ \ref{fig:filterpos}(c).
We utilized the setup in Fig. \ref{fig:prog_setup}, where we illuminated a pinhole with a 520 nm laser.
We placed a spatial mask on a horizontal translation stage to block various adjacent wavelengths.
We then measured the image of a pinhole and fit an appropriate curve to the measured PSF.
Evidently, the PSF has a larger spread as the gap between target wavelength and central wavelength of filter increases.
This is true even if the pupil code were Gaussian mask and the filter were a slit, as shown in Fig.\ \ref{fig:filterpos}(d).
Ideally we require the effective width to be independent of $\lambda_1$ and $\lambda_2$. 
This is achieved if both pupil plane and the filter have a Gaussian shape. In such a case the field,
\begin{align}
\widetilde{i}_4(x) &= \text{exp}\left\{-(x - \lambda_1f\nu_0)^2/\sigma^2\right\}\text{exp}\left\{-(x - \lambda_2f\nu_0)^2/\sigma^2\right\}\\
&= \underbrace{\text{exp}\left\{-(\lambda_1-\lambda_2)^2f^2\nu_0^2/\sigma^2\right\}}_\text{amplitude}\underbrace{\text{exp}\left\{-2(x - (\lambda_1+\lambda_2)f\nu_0/2)^2/\sigma^2\right\}}_\text{aperture shape}.
\end{align}
The output field has a spread that is independent of $\lambda_1, \lambda_2$ which results in a wavelength-independent PSF.
This is illustrated by the plot of PSFs in \ref{fig:filterpos} (e) with a gaussian aperture as well as filter shape.
Figure \ref{fig:filterpos} (f) compares PSF spread in terms of spatial standard deviation for various positions of filter, clearly illustrating the wavelength-indepedent blur arising due to Gaussian-shaped pupil-code and filter.

\section{Conclusion}
We formalized the tradeoff between spectral and spatial resolution associated with a spectrally-programmable camera of the type shown in Fig.\ \ref{fig:schematic} and stated the space-spectrum uncertainty principle.
We showed through theory, simulations and real experiments that one can finely resolve space or spectrum, but not both.
Our analysis then showed that a Gaussian-shaped aperture achieves the theoretical lower bound. 
We concluded our analysis by showing that a Gaussian-shaped narrowband filter introduces a wavelength-independent spatial blur.

\section*{Acknowledgments}
The authors acknowledge support via the National Geospatial-Intelligence Agency's Academic Research Program (Award No. HM0476-17-1-2000), the NSF CAREER grant CCF-1652569 and the NSF Expeditions award 1730147. Vishwanath Saragadam also gratefully acknowledges support via the Prabhu and Poonam Goel fellowship.

\bibliography{refs}

\begin{thebibliography}{10}
\newcommand{\enquote}[1]{``#1''}

\bibitem{cloutis1996review}
E.~Cloutis, \enquote{Review article hyperspectral geological remote sensing:
  Evaluation of analytical techniques,} {\protect\JournalTitle{International J.
  Remote Sensing}} \textbf{17}, 2215--2242 (1996).

\bibitem{colthup2012introduction}
N.~Colthup, \emph{Introduction to Infrared and Raman Spectroscopy} (Elsevier,
  2012).

\bibitem{lichtman2005fluorescence}
J.~W. Lichtman and J.-A. Conchello, \enquote{Fluorescence microscopy,}
  {\protect\JournalTitle{Nature Methods}} \textbf{2}, 910:1--11 (2005).

\bibitem{saragadam2019programmable}
V.~Saragadam and A.~C. Sankaranarayanan, \enquote{Programmable
  spectrometry--per-pixel classification of materials using learned spectral
  filters,} {\protect\JournalTitle{arXiv preprint arXiv:1905.04815}}  (2019).

\bibitem{zhi2019multispectral}
T.~Zhi, B.~R. Pires, M.~Hebert, and S.~G. Narasimhan, \enquote{Multispectral
  imaging for fine-grained recognition of powders on complex backgrounds,} in
  \emph{IEEE Intl. Conf. Comp. Vision and Pattern Recognition (CVPR),}  (2019).

\bibitem{mohan2008agile}
A.~Mohan, R.~Raskar, and J.~Tumblin, \enquote{Agile spectrum imaging:
  Programmable wavelength modulation for cameras and projectors,} in
  \emph{Comp. Graphics Forum,}  (2008).

\bibitem{love2014full}
S.~P. Love and D.~L. Graff, \enquote{Full-frame programmable spectral filters
  based on micromirror arrays,} {\protect\JournalTitle{J.
  Micro/Nanolithography, MEMS, and MOEMS}} \textbf{13}, 1--11 (2014).

\bibitem{lin2014dual}
X.~Lin, G.~Wetzstein, Y.~Liu, and Q.~Dai, \enquote{Dual-coded compressive
  hyperspectral imaging,} {\protect\JournalTitle{Optics Letters}} \textbf{39},
  2044--2047 (2014).

\bibitem{saragadam2018krism}
V.~Saragadam and A.~Sankaranarayanan, \enquote{{KRISM}---krylov subspace-based
  optical computing of hyperspectral images,} {\protect\JournalTitle{ACM Trans.
  Graphics}} \textbf{38}, 148:1--14 (2019).

\bibitem{august2013compressive}
Y.~August, C.~Vachman, Y.~Rivenson, and A.~Stern, \enquote{Compressive
  hyperspectral imaging by random separable projections in both the spatial and
  the spectral domains,} {\protect\JournalTitle{Appl. Optics}} \textbf{52},
  D46--D54 (2013).

\bibitem{wagadarikar2008single}
A.~Wagadarikar, R.~John, R.~Willett, and D.~Brady, \enquote{Single disperser
  design for coded aperture snapshot spectral imaging,}
  {\protect\JournalTitle{Appl. Optics}} \textbf{47}, B44--B51 (2008).

\bibitem{kittle2010multiframe}
D.~Kittle, K.~Choi, A.~Wagadarikar, and D.~J. Brady, \enquote{Multiframe image
  estimation for coded aperture snapshot spectral imagers,}
  {\protect\JournalTitle{Appl. Optics}} \textbf{49}, 6824--6833 (2010).

\bibitem{lctfwiki}
Wikipedia, \enquote{Liquid crystal tunable filter,}
  \url{https://en.wikipedia.org/wiki/Liquid_crystal_tunable_filter} (2019).
  [Online; accessed: 2019-07-18].

\bibitem{GRAMI201641}
A.~Grami, \enquote{Chapter 3 - signals, systems, and spectral analysis,} in
  \emph{Introduction to Digital Communications,}  (Academic Press, 2016), pp.
  41 -- 150.

\bibitem{goodman2005introduction}
J.~W. Goodman, \emph{Introduction to Fourier optics} (Roberts and Company
  Publishers, 2005).

\bibitem{opticalphysics1998}
S.~G. Lipson, H.~Lipson, and D.~S. Tannhauser, \emph{Optical Physics}
  (Cambridge University Press, 1995).

\end{thebibliography}

\begin{appendices}
	\section{Spatio-spectral Blur due to Pupil Code}\label{section:appendix1}
	We note that all our analysis is for spatially-incoherent light; any phase component is hence irrelevant.
	We rely on Fourier transform property of a thin lens \cite{goodman2005introduction} as well as the derivation in \cite{saragadam2018krism}.
	Assume that the complex field distribution on plane P1 is $i_1(x, y, \lambda)$.
	Then the field distribution on P2 that is $2f$ away is given by the scaled Fourier transform relationship, $i_2(x_2, y_2, \lambda) = \frac{1}{j\lambda f}I_1\left(\frac{x_2}{\lambda f}, \frac{y_2}{\lambda f}, \lambda\right)$, where $I_1(u, v)$ is the Fourier transform of $i_1(x, y)$.
	Propagating the signal through the optical setup simply requires us to perform such operations iteratively.
	
	Consider a single spatial point on P1 of the form $i_1(x_1, y_1, \lambda) = s(x_0, y_0, \lambda)\delta(x_1 - x_0, y_1 - y_0)$.
	Any arbitrary image can then be treated as infinite such point sources.
	The intensity distribution on plane P2 is \[i_2(x_2, y_2, \lambda) = \frac{1}{j\lambda f}s(x_0, y_0, \lambda) \text{exp}\left\{-\frac{2\pi j}{\lambda f}(x_0 x_2 + y_0 y_2)\right\}.\]
	Let $a(x, y)$ be the complex amplitude of the pupil code placed on P2. Then the intensity just after the aperture is given by,
	\begin{equation}
	\widehat{i}_2(x_2, y_2, \lambda) = \frac{1}{j\lambda f}s(x_0, y_0, \lambda) \text{exp}\left\{-\frac{2\pi j}{\lambda f}(x_0 x_2 + y_0 y_2)\right\}\times a(x_2, y_2)
	\end{equation}
	With a similar derivation, we can show that the field distribution on P3 just before the diffraction grating is,
	\begin{align}
	i_3(x_3, y_3, \lambda) = \frac{1}{j\lambda f}\widehat{I}_2\left(\frac{x_3}{\lambda f}, \frac{y_3}{\lambda f}\right)
	=\frac{1}{(j\lambda f)^2}s(x_0, y_0, \lambda) A\left(\frac{x_3 + x_0}{\lambda f}, \frac{y_3 + y_0}{\lambda f}\right),
	\end{align}
	where $A(u, v)$ is the Fourier transform of $a(x, y)$.
	For simplicity of analysis, we consider the diffraction grating to be a series of narrow band slits, modeled as an impulse train along $x$-axis as,
	\begin{equation}
	d(x, y) = \sum_{k=-\infty}^{\infty} \delta\left(x - \frac{k}{\nu_0}\right),
	\end{equation}
	where $\nu_0$ is the groove density. Then the propagated field just after the grating is given by,
	\begin{align}
	\widehat{i}_3(x_3, y_3, \lambda) &= i_3(x_3, y_3, \lambda) d(x_3, y_3)\\
	&= \frac{1}{(j\lambda f)^2}s(x_0, y_0, \lambda) A\left(\frac{x_3 + x_0}{\lambda f}, \frac{y_3 + y_0}{\lambda f}\right) \times \sum_{k=-\infty}^{\infty} \delta\left(x_3 - \frac{k}{\nu_0}\right)
	\end{align}
	
	Using Fourier transform property of lens again, the field on P4 is,
	\begin{align}
	i_4(x_4, y_4, \lambda) &= \frac{1}{j\lambda f}\widehat{I}_3\left(\frac{x_4}{\lambda f}, \frac{y_4}{\lambda f}\right)
	= \frac{1}{j\lambda f}\left(\frac{1}{j\lambda f}\right)^2I_3\left(\frac{x_4}{\lambda f}, \frac{y_4}{\lambda f}\right) \ast D\left(\frac{x_4}{\lambda f}, \frac{y_4}{\lambda f}\right)\\ 
	&= \frac{1}{j\lambda f}\left(\frac{1}{j\lambda f}\right)^2I_3\left(\frac{x_4}{\lambda f}, \frac{y_4}{\lambda f}\right) \ast \left(\delta\left(\frac{y_4}{\lambda f}\right) \sum_{k=-\infty}^{\infty} \delta\left(\frac{x_4}{\lambda f} - k\nu_0\right)\right)\\
	&= -\frac{1}{j\lambda f}s(x_0, y_0, \lambda) \sum_{k=-\infty}^{\infty} a(-(x_4 - k\nu_0 \lambda f), -y_4)\text{exp}\left\{j\frac{2\pi}{\lambda f}(x_0(x_4 - k\lambda f \nu_0) + y_0y_4)\right\},
	\end{align}
	leading to an equation which shows multiple, spectrally-dispersed copies of the aperture $a(x, y)$ along the $x$-axis.
	Our optical setup is designed to propagate only the first order and hence we retain the $k=1$ copy, giving us,
	\begin{align}
	i_4(x_4, y_4, \lambda) &= -\frac{1}{j\lambda f}s(x_0, y_0, \lambda) \underbrace{a(-(x_4 - \nu_0 \lambda f), -y_4)}_\text{spectrally-shifted a(x, y)}\text{exp}\left\{j\frac{2\pi}{\lambda f}(x_0(x_4 - \lambda f \nu_0) + y_0y_4)\right\}.
	\end{align}
	Finally, propagating the signal one more lens away, we get,
	\begin{align}
	i_5(x, y, \lambda ) &= \frac{1}{(j\lambda f)^2}\text{exp}\left\{-j2\pi x_5 \nu_0\right\} s(x_0, y_0, \lambda) A\left(-\frac{x_5 + x_0}{\lambda f}, -\frac{y_5 + y_0}{\lambda f}\right).
	\end{align}
	
	\paragraph{Intensity measurements.}
	Consider cameras placed on planes P4 and P5 with a spectral response of $c(\lambda)$. The intensity measurement on P4,
	\begin{align}
	M_4(x_4, y_4) &= \int_{\lambda} |i_4(x_4, y_4, \lambda)|^2 c(\lambda) d\lambda\\
	&= \int_{\lambda} \frac{1}{\lambda^2 f^2} |s(x_0, y_0, \lambda)|^2 |a(-(x_4 - \nu_0 \lambda f), -y_4)|^2 c(\lambda) d\lambda\\
	&= \widehat{S}\left(x_0, y_0, \frac{x_4}{f \nu_0}\right) \ast |a(-x_4, -y_4)|^2,
	\end{align}
	where $\widehat{S}\left(x_0, y_0, \lambda\right) = \frac{1}{\lambda^2 f^2} |s(x_0, y_0, \lambda)|^2 c(\lambda)$ is the measured intensity of the scene point.
	Extending to all points $(x_0, y_0)$, we get,
	\begin{align}
	M_4(x_4, y_4) &= \int_{x_0} \int_{y_0} \widehat{S}\left(x_0, y_0, \frac{x_4}{f \nu_0}\right) \ast |a(-x_4, -y_4)|^2\\
	&= S\left(\frac{x_4}{\lambda f}\right) \ast |a(-x_4, -y_4)|^2
	\label{eq:spec_exp}.
	\end{align}
	Here, $S(\lambda)$ is an integral of spectrum of all spatial points. Equation \eqref{eq:spec_exp} shows that the aperture function $a(x, y)$ results in spectral blur at every scene point.
	Similarly, the spatial image on P5,
	\begin{align}
	M_5(x, y) &= \int_{\lambda} |i_5(x_5, y_5, \lambda)|^2 c(\lambda) d\lambda\\
	&= \frac{1}{\lambda^4 f^4} \int_{\lambda} |s(x_0, y_0, \lambda)|^2 \left|A\left(-\frac{x_5+x_0}{\lambda f}, -\frac{x_5+x_0}{\lambda f}\right)\right|^2c(\lambda)d\lambda.
	\end{align}
	Computing intensity for all $(x_0, y_0)$ gives us,
	\begin{align}
	M_5(x, y) &= \int_{x_0}\int_{y_0} \frac{1}{\lambda^4 f^4} \int_{\lambda} |s(x_0, y_0, \lambda)|^2 \left|A\left(-\frac{x_5+x_0}{\lambda f}, -\frac{x_5+x_0}{\lambda f}\right)\right|^2c(\lambda)d\lambda\\
	&= \frac{1}{\lambda^4 f^4} \int_{\lambda} \underbrace{|s(x_5, y_5, \lambda)|^2 \ast \left|A\left(-\frac{x_5}{\lambda f}, -\frac{y_5}{\lambda f}\right)\right|^2}_\text{Spatial blur}c(\lambda)d\lambda.\label{eq:spat_exp}
	\end{align}
	Equation \eqref{eq:spat_exp} shows that the pupil code $a(x, y)$ introduces a spatial blur equal to a scaled version of its power spectral density (PSD), $|A(u, v)|^2$.
	For monochromatic light source,  \eqref{eq:spat_exp} is simply a convolution of scene's image with a scaled PSD of $a(x, y)$; for polychromatic sources, this expression has a spectrally-dependent PSF, which does not follow a convolution model. To make analysis simple, we assume that the shape of the PSF is approximately the same over a small range of wavelengths. Then the resultant expression for the spatial image is, 
	\begin{align}
	M_5(x_5, y_5) &= \left(\frac{1}{\lambda^4 f^4}|s(x_5, y_5, \lambda)|^2\right) \ast \left|A\left(-\frac{x_5}{\lambda_c f}, -\frac{y_5}{\lambda_c f}\right)\right|^2\\
	&= I(x_0, y_0) \ast \left|A\left(-\frac{x_5}{\lambda_c f}, -\frac{y_5}{\lambda_c f}\right)\right|^2,
	\label{eq:spat_exp_poly}
	\end{align}
	where $I(x_0, y_0)$ is a scaled grayscale image of the scene, and $\lambda_c$ is a chosen, central wavelength of the spectral range.
	
	\paragraph{Spectral and spatial blurs.}
	For brevity and ease of understanding, we drop the $y$ axis as it does not affect the spectral blur.
	From  \eqref{eq:spec_exp} and  \eqref{eq:spat_exp_poly}, we get the following expressions for spectral and spatial blurs,
	\begin{align}
	h_\lambda (\lambda) = |a(-\lambda f \nu_0)|^2, \qquad 
	h_x (x) = \left|A\left(-\frac{x}{\lambda f}\right)\right|^2.
	\end{align}
	
	\begin{figure}[!tt]
		\centering
		\begin{subfigure}[c]{0.48\textwidth}
			\centering
			\includegraphics[width=\textwidth]{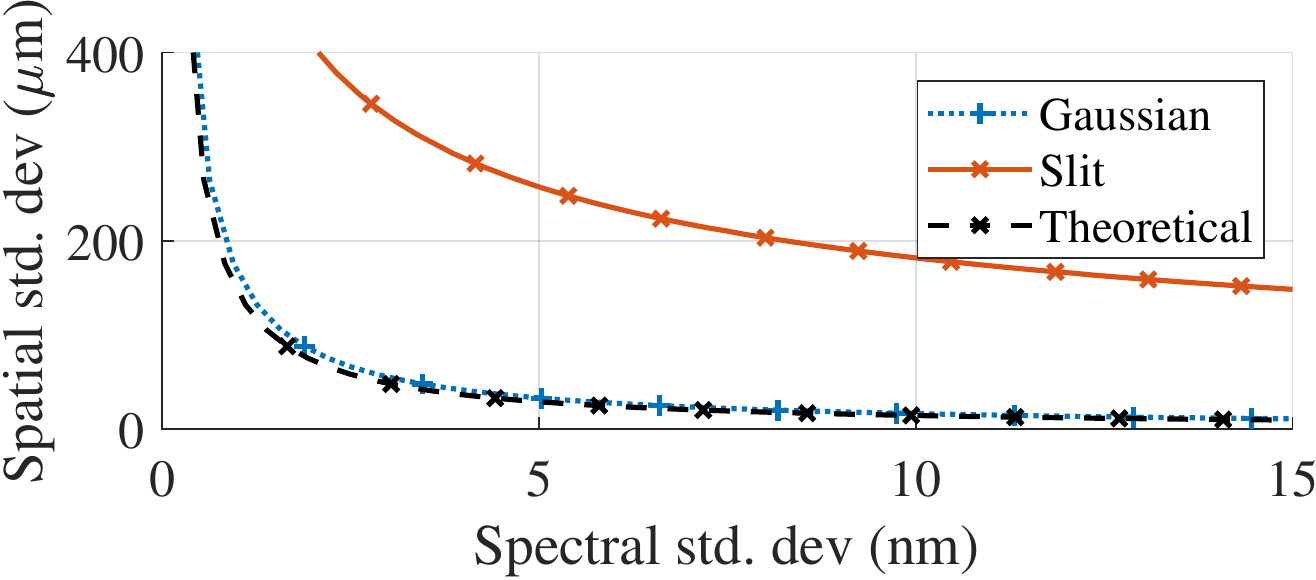}
			\caption{Standard deviation}
		\end{subfigure}
		\begin{subfigure}[c]{0.48\textwidth}
			\centering
			\includegraphics[width=\textwidth]{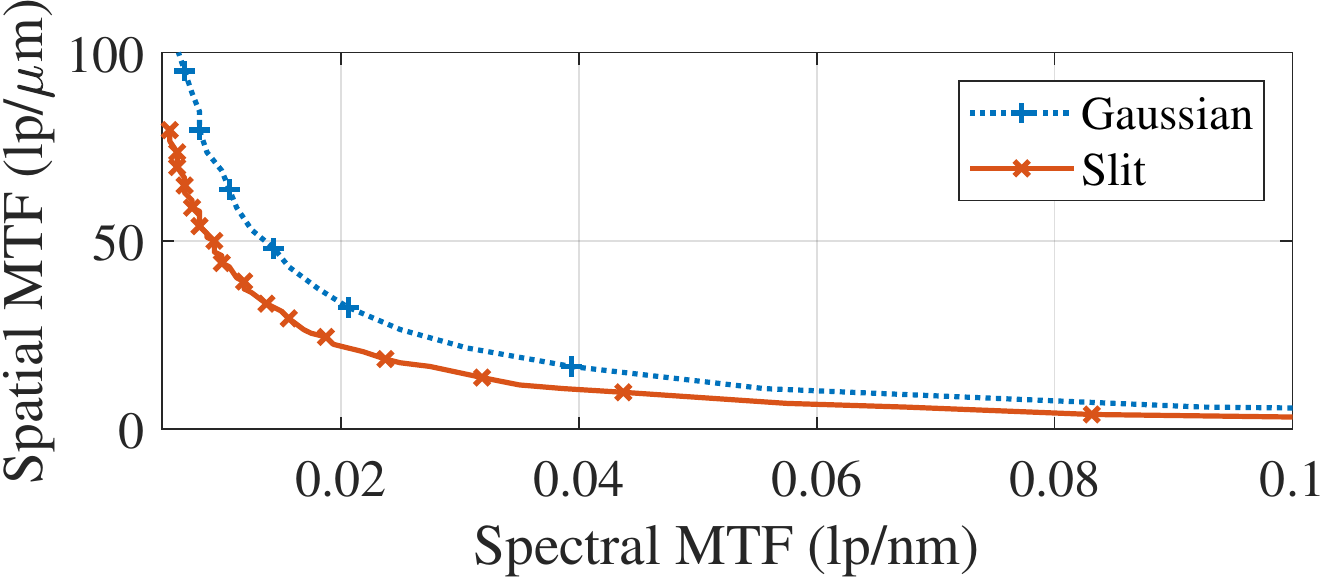}
			\caption{MTF at $30\%$ contrast ratio}
		\end{subfigure}
		\caption{\small \textbf{Simulations on common aperture shapes.} (a) compares spectral and spatial standard deviations whereas (b) shows spectral and spatial MTF at $30\%$ contrast. Gaussian codes achieve theoretical limit when resolution metric is standard deviation of window.}
		\label{fig:var_mtf}
	\end{figure}
	
	\subsection{Verification using simulations}
	We provide a validation of our theory with simulations. We specifically compared a box aperture that simulates a slit or a fully open aperture, and a Gaussian aperture.
	For the purpose of exposition, we used $f=75$ mm and a diffraction grating of $300$ groves/mm. 
	Figure \ref{fig:var_mtf}(a) shows a plot of spatial and spectral standard deviations and (b) shows a plot of spatial and spectral modulation transfer function (MTF) at $30\%$ contrast ratio. The plots show a clear trade off between the two resolutions, independent of resolution metric.
	We also observe that Gaussian codes achieve the theoretical limit for standard deviation. Hence we conclude that the space-spectrum bandwidth product is a tight bound.
\end{appendices}

\end{document}